\documentclass[aps,twocolumn,pra,tightenlines,floatfix,showpacs,superscriptaddress,longbibliography]{revtex4-1}

\usepackage{graphicx}
\usepackage{amsmath}
\usepackage{mathtools}
\usepackage{amssymb}
\usepackage{times}
\usepackage{epstopdf}
\usepackage[english]{babel}
\usepackage{natbib}
\usepackage{hyperref}
\usepackage{color}
\usepackage{cleveref}
\usepackage{tablefootnote}
\usepackage{float}
\usepackage{subcaption}
\usepackage[version=4]{mhchem} 
\usepackage{adjustbox}
\usepackage{array, booktabs, makecell}
\usepackage{colortbl}
\usepackage{siunitx}
\let\svqty\qty
\usepackage{physics}
\let\qty\svqty
\usepackage{enumitem}
\usepackage[normalem]{ulem}
\usepackage[dvipsnames]{xcolor}

\hypersetup{colorlinks=true, citecolor=blue, linkcolor=blue}
\DeclareUnicodeCharacter{0301}{a}

\newlist{steps}{enumerate}{1}
\setlist[steps, 1]{wide=0pt, label=Step \arabic*:, font=\bfseries}

\newcommand{\grayline}{\arrayrulecolor{lightgray}\hline\arrayrulecolor{black}}
\newcommand{\vertiiii}[1]{{\vert\kern-0.25ex\vert\kern-0.25ex\vert #1 
    \vert\kern-0.25ex\vert\kern-0.25ex\vert}}

\begin{document}
\title{Self-consistent Quantum Iteratively Sparsified Hamiltonian method (SQuISH): \\ A new algorithm for efficient Hamiltonian simulation and compression} 
\begin{abstract}
    It is crucial to reduce the resources required to run quantum algorithms and simulate physical systems on quantum computers due to coherence time limitations.
    With regards to Hamiltonian simulation, a significant effort has focused on building efficient algorithms using various factorizations and truncations, typically derived from the Hamiltonian alone.
     We introduce a new paradigm for improving Hamiltonian simulation and reducing the cost of ground state problems based on ideas recently developed for classical chemistry simulations.  
     The key idea is that one can find efficient ways to reduce resources needed by quantum algorithms by making use of two key pieces of information:  the Hamiltonian operator and an 
     approximate ground state wavefunction.  
     We refer to our algorithm as the \textit{Self-consistent Quantum Iteratively Sparsified Hamiltonian} (SQuISH). 
     By performing our scheme iteratively, one can drive SQuISH to create an accurate wavefunction using a truncated, resource-efficient Hamiltonian. 
     Utilizing this truncated Hamiltonian provides an approach to reduce the gate complexity of ground state calculations on quantum hardware.  
     As proof of principle, we implement SQuISH using configuration interaction for small molecules and coupled cluster for larger systems. 
     Through our combination of approaches, we demonstrate how 
     SQuISH performs on a range of systems, the largest of which would require more than 200 qubits to run on quantum hardware.
     Though our demonstrations are on a series of electronic structure problems, our approach is relatively generic and hence likely to benefit additional applications where the size of the problem Hamiltonian creates a computational bottleneck.

\end{abstract}
\author{Diana B. Chamaki}
\affiliation{Quantum Artificial Intelligence Laboratory (QuAIL), Exploration Technology Directorate,
NASA Ames Research Center, Moffett Field, CA 94035, USA}
\affiliation{USRA Research Institute for Advanced Computer Science, Mountain View, California 94043, USA}
\author{Stuart Hadfield}
\affiliation{Quantum Artificial Intelligence Laboratory (QuAIL), Exploration Technology Directorate,
NASA Ames Research Center, Moffett Field, CA 94035, USA}
\affiliation{USRA Research Institute for Advanced Computer Science, Mountain View, California 94043, USA}
\author{Katherine Klymko}
\affiliation{NERSC, Lawrence Berkeley National Laboratory, Berkeley, California 94720, USA}
\author{Bryan O'Gorman}
\affiliation{IBM Quantum, IBM T.J. Watson Research Center, Yorktown Heights, NY, USA}
\author{Norm M. Tubman}
\affiliation{Quantum Artificial Intelligence Laboratory (QuAIL), Exploration Technology Directorate,
NASA Ames Research Center, Moffett Field, CA 94035, USA}
\maketitle

\section{Introduction} 
\begin{figure*} [t]
    \centering
    \includegraphics[width=\textwidth]{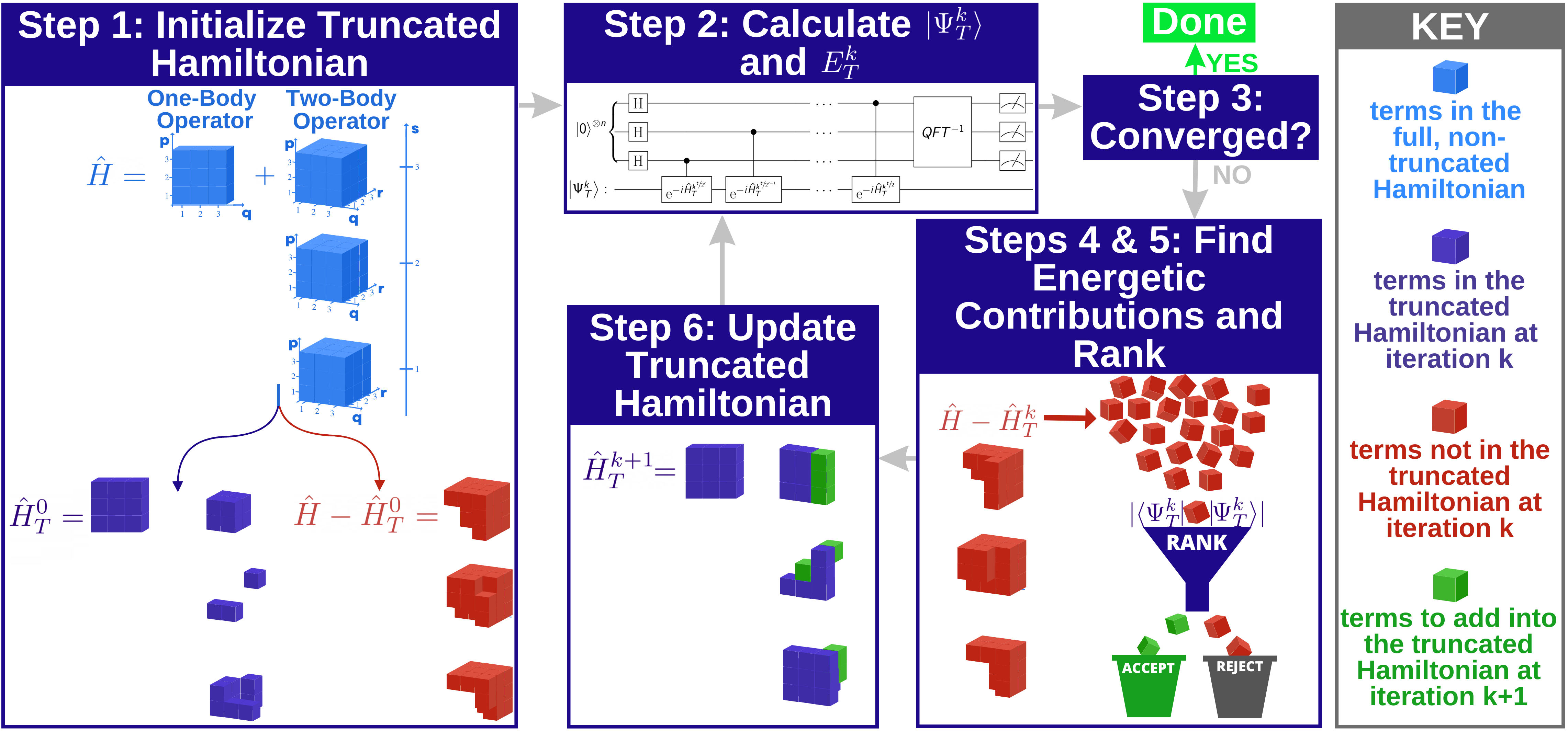}
    \caption{Schematic diagram of SQuISH (see Section~\ref{sec:SQuISH} for detailed explanation of each step). 
    The Hamiltonian is illustrated with a three dimensional representation of the 2D one-body operator and 4D two-body operator. 
    Each cube represents a single Hamiltonian term, $h_{pq}\hat{a}_p^\dagger\hat{a}_q$ or $h_{pqrs}\hat{a}_p^\dagger\hat{a}_q^\dagger \hat{a}_r \hat{a}_s$.
    This figure gives a hypothetical example of SQuISH applied to a 3 spatial orbital molecule. 
    The initial Hamiltonian (in Step 1) includes all one-body terms and a subset of the two-body terms as an example.
    QPE is shown as possible quantum algorithm used for the truncated Hamiltonian ground state calculation.}   
    \label{SC_chart}
\end{figure*}

The electronic structure problem gives insight into reaction rates, geometry of stable structures, and determination of optical properties, among many others~\cite{helgaker, McArdle2020, Bauer2020}. Classical methods such as full configuration interaction, which simulate the entire Hilbert space,  are limited to small systems due to the computational scaling with increasing orbitals and electrons.  ~\cite{Thogersen2004, aspuru-guzik2005}. While there are classical approaches to calculate such solutions approximately~\cite{szabo2012modern,helgaker2014molecular}, it is anticipated that quantum computers may offer an exponential advantage in calculating exact solutions for some quantum computational chemistry problems \cite{aspuru-guzik2005,mcardle2020quantum}.  

Quantum phase estimation (QPE) \cite{Kitaev1995, Abrams1999} is a promising quantum algorithm to solve electronic structure problems near-exactly.
QPE and its variants evolve a 
quantum state in time under a given Hamilton, which 
amounts to solving the  
Schr\"{o}dinger equation for different times. 
A few of the 
most prominent time evolution methods are Trotter-Suzuki product formulas~\cite{trotter1959product,suzuki1976generalized,lloyd1996universal,aharonov2003adiabatic,berry2007efficient,hadfield2018divide,childs2021theory}, related randomized approaches such as qDRIFT~\cite{zhang2012randomized,Campbell_2019,childs2019faster},
and post-Trotter approaches such as qubitization~\cite{low2017optimal,Low_2019}. 
 
If we consider Trotterization based methods, performing just one Trotter step for a molecule scales as $N^4$ (the number of Hamiltonian terms) if approximations on the Hamiltonian are not used. 
Here $N$ is the number of spin orbitals used for a given system, typically much larger than the number of electrons as more accurate basis sets are employed~\cite{seeley2012, hastings2014}. 
Generally, the cost of simulating a Hamiltonian using these methods depends on the number of terms as well as their magnitudes with respect to particular operator norms~\cite{Childs_2021, Loaiza_2022}.

For the most part, there is a direct correspondence between quantum circuit depth and requisite physical coherence time.
Due to interaction
with the environment, even if one has a quantum device with sufficiently many qubits, maintaining a quantum state long enough to perform calculations reliably is a significant challenge \cite{Zurek1981, Joos1985, Preskill2012}. 
With current quantum hardware, quantum states are especially short-lived. 
Thus, Hamiltonians with a large number of nonnegligible terms at each step cannot practically be implemented on such hardware.
Noisy-intermediate scale quantum (NISQ) era devices provide severe constraints in addition to other limitations, such as noise or connectivity, so we must continually improve upon the efficiency of quantum algorithms to run more effectively on current and future  hardware~\cite{preskill2018quantum} .

A series of recent works has focused on reducing the gate complexity of Hamiltonian simulation for chemistry applications~\cite{wecker2014gate,
hastings2014,
mcclean2014exploiting,
poulin2014step,
babbush2015chemical,
BabbushWiebe2018, BabbushGidney2018, LowWiebe2019, Kivlichan2020, McClean2021, Lee2021, Motta2021}. 
These papers propose methods for reducing gate complexity by
approximations that utilize the Hamiltonian alone. 
In this paper we propose a significant change in the truncation/factorization process, noting that the Hamiltonian alone does not contain all the information needed to assess which terms are important for 
particular applications.   
Hamiltonians contain many eigenstates of interest, and the importance of the terms in the Hamiltonian are dependent on which eigenstates are under study for a given application.  
In other words, by ignoring eigenstate data in any resource reduction approach, we are potentially leaving out critical information.

Here, we present the \textit{Self-consistent Quantum Iteratively Sparsified Hamiltonian} (SQuISH) algorithm, which utilizes both the Hamiltonian and the (approximate) eigenstate of interest to produce a truncated Hamiltonian with reduced complexity. 
We note that SQuISH is inspired by the Adaptive Sampling Configuration Interaction (ASCI) classical algorithm~\cite{Tubman2016,Tubman2018, Tubman2020} but with swapped roles of the Hamiltonian and approximate wavefunction (see Appendix~\ref{sec:ASCI} for more detail on ASCI). 
It can also be used in combination with other recently developed compression techniques to build an even more compact Hamiltonian.

The structure of the paper is as follows: Section \ref{sec:theoretical background} provides the theoretical background.
Section~\ref{sec:SQuISH} gives a detailed description of SQuISH. 
Section~\ref{sec: SQuISH extensions} discusses potential variants of SQuISH.
Section~\ref{sec:simulation details} explains the details of our implementation. 
Section~\ref{sec:results} demonstrates the quantum resource advantages SQuISH provides using coupled cluster and configuration interaction for various molecules in a range of different bases. 
Section~\ref{sec:conclusion} contains concluding remarks.

\section{Theoretical Background} \label{sec:theoretical background}

The theoretical background used in this work is given in this section. 
We define the second quantized electronic Hamiltonian and reduced density matrices (RDMs) in Section~\ref{sec:Electronic Hamiltonian}. 
Then, in  Section~\ref{sec: quantum algorithms}, we discuss the basics of two quantum algorithms we use.

\subsection{The Electronic Hamiltonian} \label{sec:Electronic Hamiltonian}

In this work, we consider the second quantized electronic Hamiltonian with clamped nuclei \cite{helgaker,szabo2012modern} defined in terms of fermionic creation and annihilation operators $\hat{a}^\dagger$, $\hat{a}$
as 
\begin{equation} \label{2ndquantizedham}
\begin{split}
    &\hat{H} = \sum_{\sigma \in \{\uparrow, \downarrow\}}\sum_{p,q}^M h_{pq}\hat{a}_{p,\sigma}^\dagger \hat{a}_{q,\sigma} \\ 
    &+ \frac{1}{2}\sum_{\alpha, \beta \in \{\uparrow, \downarrow\}}\sum_{p,q,r,s}^M h_{pqrs}\hat{a}_{p,\alpha}^\dagger \hat{a}_{q,\beta}^\dagger \hat{a}_{r, \alpha} \hat{a}_{s, \beta} + h_{nuc}, 
\end{split}
\end{equation}
where in atomic units we have
\begin{equation}
    h_{pq} = \int \phi_{p}^*(\mathbf{r}) \left(-\frac{1}{2}\nabla^2 - \sum_I \frac{Z_I}{r_I}\right)\phi_{q}(\mathbf{r}) d\mathbf{r},
\end{equation}

\begin{equation}
    h_{pqrs} = \int \int \frac{\phi_{p}^*(\mathbf{r}_1) \phi_{q}^*(\mathbf{r}_2) \phi_{r}(\mathbf{r}_1) \phi_{s}(\mathbf{r}_2)}
    {r_{12}}
    d\mathbf{r}_1 d\mathbf{r}_2,
\end{equation}

\begin{equation}
    h_{\text{nuc}} = \frac{1}{2} \sum_{I \neq J}\frac{Z_I Z_J}{R_{IJ}}.
\end{equation}
In the equations above $M$ is the number of spacial orbitals, $N=2M$ is the number of spin orbitals, the set $\{\phi_i\}_i^M$ is a basis of $M$ orthonormal spatial orbitals, $Z_I$ are the nuclear charges, $r_I$ are the electron-nuclear separation distances, $r_{12}$ is the electron-electron separation, $R_{IJ}$ are the internuclear separations, and $h_{nuc}$ is the nuclear repulsion energy.

A finite basis approximation must be used to realize the Hamiltonian in practice. 
The one- and two-body operators are used to calculate the Hamiltonian coefficients. 
Then to solve the electronic structure problem on a quantum computer (i.e., estimate the ground state energy), each term in the Hamiltonian gets suitably mapped onto qubit operators, for which a variety of mappings have been proposed~\cite{whitfield2011simulation,seeley2012bravyi,setia2019superfast}. 
Generally, the more terms there are in the Hamiltonian, the larger the gate complexity required for a simulation or energy measurement.

We will use reduced density matrices (RDMs) and the spin free electron repulsion integrals (ERIs) to calculate the ground state energy of the Hamiltonian and expectation values of individual terms. 
Moving forward, the spin indices ($\sigma, \alpha, \beta$) are suppressed. 
We denote the elements of the spin free 1RDM as $\gamma_{p,q}$ and the 2RDM as $\Gamma_{p,q,r,s}$ defined as follows:
\begin{align} \label{1rdm}
    \gamma_{p,q} &= \expval{\hat{a}_p^\dagger\hat{a}_q}{\psi}\\
    \Gamma_{p,q,r,s} &= \expval{\hat{a}_p^\dagger\hat{a}_q^\dagger\hat{a}_r \hat{a}_s}{\psi}.\label{2rdm}
\end{align}

\subsection{Relevant Quantum Algorithms} \label{sec: quantum algorithms}

In addition to QPE (which we introduced in the introduction), the variational quantum eigensolver (VQE)~\cite{Peruzzo2014, McClean2016} is a prominent paradigm in solving for ground and low-lying energy eigenstates.
VQE is an iterative quantum-classical hybrid algorithm that repeatedly prepares and measures parameterized quantum circuits built from a particular ansatz to obtain approximations of a target energy and its corresponding eigenstate.
The canonical QPE algorithm takes a qubit register containing an approximate eigenstate, together with an ancillary register for storing the eigenvalue that is initialized in superposition by applying a Hadamard gate to each ancilla qubit, applies a sequence of controlled unitary gates, then applies the inverse quantum Fourier transformation to the ancilla register, and finally measures the ancilla qubits in the computational basis to obtain the desired energy estimate.

VQE is particularly promising in the current era of NISQ hardware because it is often compatible with relatively short coherence times. 
However, there are numerous challenges associated with practically implementing VQE, especially when applied to increasingly larger systems~\cite{cerezo2021variational,tilly2021variational,bharti2022noisy}. 
For example, solving the classical parameter optimization problem in some cases becomes impractical as the cost landscape becomes more complex with the growth in the number of parameters~\cite{Huembeli2021,bittel2021training,mcclean2018barren,wang2021noise}. 
Additionally, VQE generally requires a large number of repeated state preparations and measurements to evaluate the required Hamiltonian expectation values~\cite{izmaylov2019revising,huggins2021efficient}. 
QPE, on the other hand, provides near-exact solutions without the aforementioned problems, but is challenging to implement due to long circuit depths \cite{Mohammadbagherpoor_2019,Mohammadbagherpoor_2019_ieee}. 

\section{Self-consistent, Iterative Truncation} \label{sec:SQuISH} 
For electronic structure problems, the number of terms in the Hamiltonian scales with the number of orbitals, like $N^4$, as is easily seen from the quartic terms in ~Eqn.~\eqref{2ndquantizedham}. Such a calculation is challenging to execute on quantum hardware. To our knowledge, one of the largest demonstrations of an accurate quantum simulation has been demonstrated for only 6 orbitals (12 qubits) \cite{google12qubits2020}.
Thus a strategy of aggressively truncating the number of terms in the Hamiltonian while achieving a targeted accuracy is a 
highly desirable, especially for NISQ applications. 
Chemical accuracy is a standard target in determining molecular energies, but less restrictive goals are sometimes also of interest for particular applications. SQuISH may be a useful tool in achieving such goals. In this section, we fully detail each component of SQuISH.

\subsection{Self-consistent Quantum Iteratively Sparsified Hamiltonian Algorithm} \label{self-consistent}
Moving forward, we will use the variable definitions given in Table~\ref{tab:variables}.
Fig.~\ref{SC_chart} provides a visual representation of SQuISH applied to a hypothetical system with 3 spatial orbitals.
Each step labeled in Fig.~\ref{SC_chart} corresponds to the steps detailed below. 

\begin{table}[H]
\begin{center}
{\renewcommand{\arraystretch}{1.25}
\setlength{\tabcolsep}{0.65em}
\begin{tabular}{ l  l }
\hline
Variable & Description\\ \hline
 $\hat{H}$ & Electronic Hamiltonian \\ \grayline
 $\hat{H}_{T}^k$ & Truncated Hamiltonian at iteration $k$ \\  \grayline
 $\ket{\Psi_0}$ & Ground state wavefunction of  $\hat{H}$ \\   \grayline
 $\ket{\Psi_{T, 0}^k}$ & Ground state wavefunction of  $\hat{H}_{T}^k$ \\\grayline
 $E_0$ & Accurate ground state energy, i.e. $\expval{\hat{H}}{\Psi_0}$ \\  \grayline
 $E_{T}^k$ & Ground state energy of $\hat{H}_{T}^k$\\ \grayline
 $\mathbb{I}^k$ & Set of one- and two-body indices in $\hat{H}_{T}^k$ at iteration $k$\\ \grayline
 $\mathbb{T}^k$ & Set of one- and two-body indices in $\hat{H}-\hat{H}_{T}^k$ \\ & at iteration $k$ \\ \grayline
 $\delta$ & Convergence parameter\\ \grayline
 $\eta$ & Number of electrons\\ \grayline
 $l$ & Number of observables\\ \hline
\end{tabular}}
\caption{\label{tab:variables} The variables and corresponding definitions used in this section and future sections.}
\end{center}
\end{table}

\begin{steps}
    \item Define the truncated Hamiltonian at iteration $k = 0$ with a small number of Hamiltonian terms, which we call the base terms. 
    
    Select $\mathbb{I}^0$ such that it includes at least all terms associated with the Hartree Fock ground state. 
    
    The initial truncated Hamiltonian is then 
    \begin{equation}
    \hat{H}_{T}^0 = \sum_{p,q \in \mathbb{I}^0} h_{pq}\hat{a}_p^\dagger\hat{a}_q+ \frac{1}{2} \sum_{p,q,r,s \in \mathbb{I}^0} h_{pqrs}\hat{a}_p^\dagger\hat{a}_q^\dagger\hat{a}_r\hat{a}_s.
    \end{equation}
    
    \textbf{Repeat the following steps until convergence}:
    \item Find the ground state wavefunction and energy of $\hat{H}_{T}^k$.  
   
    \item Check for convergence ($k>1$): 
    Calculate the difference between the approximate energy from the current and previous iteration. 
    
    SQuISH terminates when $\abs{E_T^{k-1} - E_T^k} < \delta$, or if $\mathbb{T}^k=\emptyset$.
     
    \item Calculate energetic contributions of each term in $\mathbb{T}^k$.

    Find the expectation value of terms using 
    \begin{equation} \label{term_expval}
    \left\{\epsilon_{i} =
    \begin{cases}
        \expval{h_{pq} \hat{a}_p^\dagger \hat{a}_q + h.c.}{\Psi_{T,0}^k}\\
         \expval{h_{pqrs} \hat{a}_p^\dagger \hat{a}_q^\dagger \hat{a}_r \hat{a}_s + h.c.}{\Psi_{T,0}^k}
    \end{cases}
    \right\},
    \end{equation}
    
    where $h.c.$ denotes the hermitian conjugate.

    \item Rank the terms such that $\abs{\epsilon_{i}} \geq \abs{\epsilon_{{i+1}}}$ for any $i \in \{1, \dots, l-1\}$, where $l$ is the number of terms in $\mathbb{T}^k$. 

    \item Update $\hat{H}_T^{k+1}$ with the $m$ most important terms (and hermitian conjugates), and increment $k$.

Here, $m$ indicates the number of terms we want to add to the truncated Hamiltonian at the current iteration. 
Depending on the system, $m$ can be fixed or incremented over the iterations.

\textbf{End of algorithm.}
\end{steps}

Note that while for simplicity, we define SQuISH here in terms of exact eigenstates and energies, in practice, these may be computationally expensive, even for the truncated Hamiltonians, and so suitable approximations may be employed instead. We elaborate on such approximations and give several examples in subsequent sections.

\subsection{Energetic Ranking Scheme} \label{ranking_subsection}

Our ranking scheme, used in Step 4 and Step 5, is a key component in the novelty of SQuISH.
We propose a method to determine the relative importance of each term in the Hamiltonian by utilizing an iteratively updated ground state wavefunction.
Existing alternative approaches often truncate the Hamiltonian naively by excluding or transforming terms based on the value of the associated coefficients without taking advantage of any information derived from the target state of interest. 
Consider the ground state energy given in Eqn.~\eqref{ground state energy}, 
\begin{align} \label{ground state energy}
E_0 &= \expval{\hat{H}}{\Psi_0} \\
&= \sum_{p,q} \expval{h_{pq} \hat{a}_{p}^\dagger \hat{a}_{q}}{\Psi_0} \nonumber  \\ &+ \frac{1}{2}\sum_{p,q,r,s} \expval{h_{pqrs} \hat{a}_{p}^\dagger \hat{a}_{q}^\dagger \hat{a}_{r} \hat{a}_{s}}{\Psi_0}. \nonumber
\end{align}
The expectation value of each Hamiltonian term reveals how much the term contributes to the energy.  
Clearly, the naive truncation approach based on the coefficients $h_{pq}$, $h_{pqrs}$ alone does not take advantage of any information available regarding the ground state, which may be significant.  
To illustrate this point even further, the energetic contribution of each individual term in the above equation may dramatically change if we were to replace $\Psi_0$  with an excited state wavefunction. 

Thus the  ranking scheme we suggest is based on Eqn.~\eqref{ground state energy}, where we seek to \emph{approximately} include the influence of the ground state wavefunction, even before we have precise knowledge of it, rather than based on the coefficients $h_{pq}$, $h_{pqrs}$ alone. 
To proceed, we begin with a suitably truncated Hamiltonian $\hat{H}_{T}$ and corresponding approximate ground state, $\Psi_{T,0}$, which we iteratively update to calculate the energetic contribution of each term at each step. 
We determine the importance of terms based Eqn.~\eqref{term_expval}.

The iterative aspect of the algorithm is needed since we do not have the actual ground state wavefunction to perform this ranking.
Additional terms are added to the Hamiltonian at each iteration. 
If this is done in a methodical manner, the current ground state wavefunction gets closer and closer to the ground state of the full Hamiltonian of interest with each step.

\subsection{Benchmark with the Truncated Hamiltonian}\label{benchmark}

Here we describe a benchmark comparison we developed for SQuISH, which we refer to as Benchmark 1.
We consider cases in which we can easily calculate $\ket{\Psi_0}$ and can use it to rank terms (in Step 4) in place of $\ket{\Psi_{T,0}^k}$. 
When using the exact wavefunction in the ranking, we are effectively inserting each term in order according to the absolute value of its energetic contribution.
This is in comparison to the actual algorithm in which we only have an approximate trial wavefunction.

After using the exact wavefunction in the ranking, we can test the set of terms in Step 2 by using the ground state wavefunction of the truncated Hamiltonian to calculate the following energy:
\begin{equation}\label{nv_energy}
    E_{T, \text{ nv}}^k = 
    \expval{\hat{H}_T^k}{\Psi_{T,0}^k}. 
\end{equation}
We note that the energy here is non-variational with regards to the ground state energy of the full Hamiltonian.

\subsection{Improved Benchmarking with the Full Hamiltonian}\label{improved_benchmark}    

\begin{figure*} [t]
    \centering
    \includegraphics[width=\textwidth]{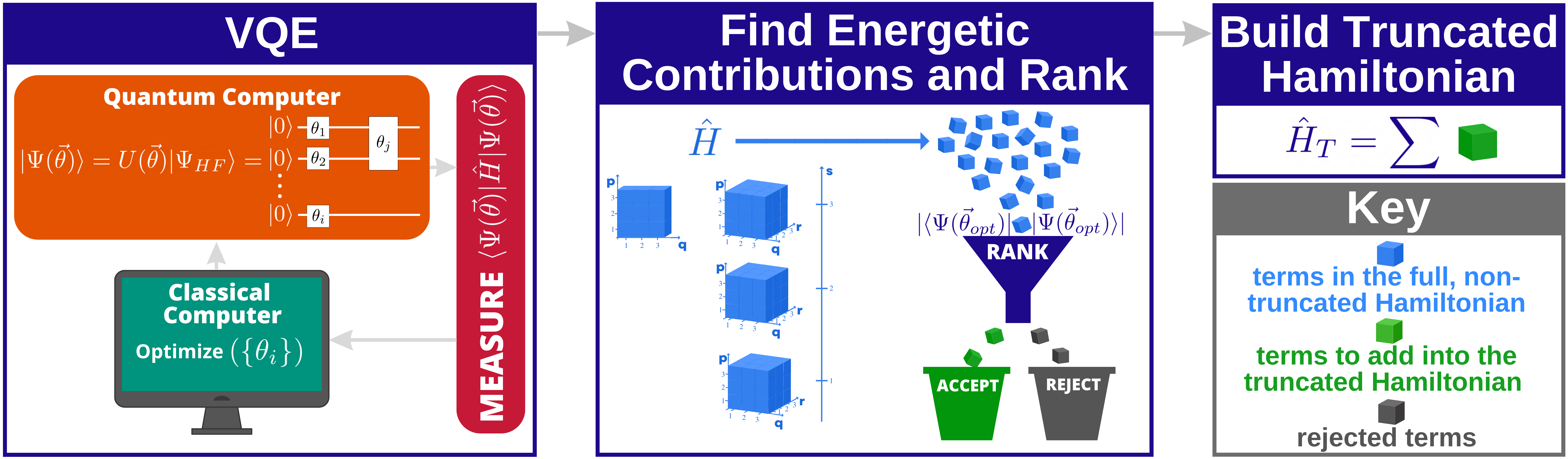}
    \caption{A schematic diagram of the non-iterative truncation with VQE using the energetic ranking scheme. 
    Each cube represents one Hamiltonian term.
    The first panel shows the calculation of the optimized ansatz, $\Psi(\vec{\theta}_{opt})$, using VQE.
    The next panel illustrates the ranking of the terms using $\Psi(\vec{\theta}_{opt})$. The last panel shows the truncated Hamiltonian as a sum of the most important terms.}
    \label{vqe_trunc_chart}
\end{figure*}

While Benchmark 1 is useful, it can be improved in practice by making use of the variational principle. 
We refer to this benchmark as Benchmark 2.
Without changing the ranking in Step 4, we can calculate the following energy:
\begin{equation}\label{v_energy}
    E_{T, \text{ v}}^k = \expval{\hat{H}}{\Psi_{T,0}^k}.  
\end{equation}
We expect a faster convergence when Eqn.~\eqref{v_energy} is used in Step 2 since it uses the full Hamiltonian to calculate the expectation value. 
However, there is a trade-off between accuracy and the number of measurements one would need to make on a quantum computer when using the full Hamiltonian in place of the truncated Hamiltonian to calculate the energy.

\subsection{Energy estimates using classical shadows} \label{sec:shadow tomography}

In each iteration $k$ of SQuISH, we have a state $\ket{\Psi_{T,0}^k}$ and need estimates of both the total energy $E_T^k$ (using the current truncated Hamiltonian) and the energetic contributions of the remaining one- and two-body terms (indexed by $\mathbb T^k$).
These can be calculated from estimates of all of the 2-RDMs.
The state of the art for estimating all 2-RDMs is based on classical shadows.
In the general framework for partial tomography using classical shadows, a Positive Operator Valued Measure (POVM) is chosen such that $l$ observables ${\left(\hat{O}_i\right)}_{i=1}^l$ from a given family can be estimated to within $\epsilon$ additive error using $O\left(\log l \max_i {\left\|{\hat{O}_i}\right\|}_{\mathrm{shadow}} / \epsilon^2\right)$ of the POVM~\cite{huang2020predicting,elben2022randomized}, where ${\left\|\cdot\right\|}_{\mathrm{shadow}}$ is a POVM-specific norm related to the variance of the estimator.
The POVM is typically specified using an ensemble of unitaries such that the POVM results from choosing a unitary uniformly at random from the ensemble and then measuring in the computational basis.
The choice of POVM depends on the type of observable to be estimated, but not on the specific observables once the type is fixed.
Several recent works have developed protocols for measuring fermionic observables using classical shadows~\cite{zhao2021fermionic,wan2022matchgate,ogorman2022fermionic,low2022classical}.
In particular, Low showed that when the state to be measured has a fixed number $\eta$ of electrons, all of the 2-RDMs can be estimated to within additive error $\epsilon$ using only

\begin{align}
\binom{\eta}{2} 
{\left(1 - \frac{\eta - 2}{N}\right)}^2 \frac{N + 1}{N - 1} \frac{1}{\epsilon^2}
\end{align}
measurements~\cite{low2022classical}.
The ensemble of unitaries used is simply Haar-random single-particle basis changes, which can be implemented in linear depth with linear connectivity (when using the Jordan-Wigner encoding)~\cite{jiang2018quantum}.
Computing the estimates of the RDMs given the sampled unitaries and measurement outcomes can be done efficiently.
Furthermore, because the measurements are non-adaptive, the estimation procedure can be trivially parallelized.

In theory, there may be a more efficient measurement protocol for the observables of interest (the total energy and energetic contributions) that exploits some structure in the observables. 
However, because we are focused on sparse truncations of the Hamiltonian terms, the excluded terms indexed by $\mathbb T^k$ can include almost all of the terms, and so we essentially need to measure all of the 2-RDMs anyway.
That being said, each energetic contribution has a coefficient $h_{pqrs}$ that is not taken into account in the scheme based on classical shadows; there is potential there for improving the measurement efficiency, which we leave to future work.

\section{Extensions of SQuISH} \label{sec: SQuISH extensions}

By adopting and expanding upon some of the methods proposed in Section~\ref{sec:SQuISH}, one can develop useful variations of SQuISH.
There are a multitude of SQuISH extensions, which can be explored in future work, but here we will discuss two such extensions. 
While we largely focus on Hamiltonian truncation for ground state problems in this paper, SQuISH can also be employed for excited state calculations by slightly altering our algorithm. 
We call this modified algorithm "multi-reference SQuISH" and explain the details. 
Additionally, our energetic ranking scheme without the iterative portion of SQuISH may be of interest. We discuss a method of using an approximate wavefunction to rank terms and truncate a Hamiltonian non-iteratively.

\subsection{Multi-reference SQuISH} \label{sec: multi-reference SQuISH}
If one is interested in truncating a Hamiltonian for dynamics, considering excited states might be useful to create a selected energy space for the dynamics and can be included in the truncation process.
We can create a multi-reference version of SQuISH such that it includes terms based on both ground and excited state energetic contributions.
In Step 2, we can calculate the $J$ truncated Hamiltonian wavefunctions of interest, such as $\{\ket{\Psi_{T,0}^k}, \ket{\Psi_{T,1}^k}, \ket{\Psi_{T,2}^k}, \dots \ket{\Psi_{T,J}^k}\}$. 
Then in Step 4, we would calculate the energetic contributions of the terms with each of the $J$ wavefunctions as follows:
\begin{equation}
\begin{gathered}
    \left\{\epsilon_{i,0} =
    \begin{cases}
        \expval{h_{pq} \hat{a}_p^\dagger \hat{a}_q + h.c.}{\Psi_{T,0}^k}\\
         \expval{h_{pqrs} \hat{a}_p^\dagger \hat{a}_q^\dagger \hat{a}_r \hat{a}_s + h.c.}{\Psi_{T,0}^k}
    \end{cases}
    \right\} \\
   \left\{\epsilon_{i,1} =
    \begin{cases}
        \expval{h_{pq} \hat{a}_p^\dagger \hat{a}_q + h.c.}{\Psi_{T,1}^k}\\
         \expval{h_{pqrs} \hat{a}_p^\dagger \hat{a}_q^\dagger \hat{a}_r \hat{a}_s + h.c.}{\Psi_{T,1}^k}
    \end{cases}
    \right\} \\
    \left\{\epsilon_{i,2} =
    \begin{cases}
        \expval{h_{pq} \hat{a}_p^\dagger \hat{a}_q + h.c.}{\Psi_{T,2}^k}\\
         \expval{h_{pqrs} \hat{a}_p^\dagger \hat{a}_q^\dagger \hat{a}_r \hat{a}_s + h.c.}{\Psi_{T,2}^k}
    \end{cases}
    \right\} \\
    \vdots\\
    \left\{\epsilon_{i,J} =
    \begin{cases}
        \expval{h_{pq} \hat{a}_p^\dagger \hat{a}_q + h.c.}{\Psi_{T,J}^k}\\
         \expval{h_{pqrs} \hat{a}_p^\dagger \hat{a}_q^\dagger \hat{a}_r \hat{a}_s + h.c.}{\Psi_{T,J}^k}
    \end{cases}
    \right\}.
\end{gathered}
\end{equation}
After ranking each set of energetic contributions in Step 5, we can update the Hamiltonian with $m$ terms consisting of the $m/J$ most important terms from each set.  Convergence criteria can be extended to include the convergence of the excited states in the loss function.

\subsection{Non-Iterative Truncation with VQE}
\label{sec:non-iterative truncation}
Algorithms like VQE can be used in combination with the energetic ranking scheme to truncate the Hamiltonian non-iteratively. 
SQuISH is reliant on an iteratively updated trial wavefunction to rank terms. 
However, a one-shot approach for the compression process may be useful in practice.
We can build the truncated Hamiltonian all in one iteration by using a sufficiently accurate wavefunction to rank all terms. 
One can use VQE to optimize a Unitary Coupled Cluster (UCC) ansatz \cite{Bartlett1989, Hoffmann1988}, a Hamiltonian variational ansatz (HVA)~\cite{Wecker2015}, or a hardware-efficient ansatz \cite{Kandala2017} in place of the iteratively updated SQuISH trial wavefunction to determine which Hamiltonian terms to exclude from the truncated Hamiltonian. 
A schematic diagram of this process is shown in Fig.~\ref{vqe_trunc_chart}.

A non-iteratively truncated Hamiltonian may be utilized both to study dynamics and find more accurate ground state solutions.
We expect that in some cases an optimized trial VQE wavefunctions generated without necessarily reaching chemical accuracy can be sufficiently good for finding the energetic contributions of terms and ranking.
Once the truncated Hamiltonian is built, one can use QPE or other approaches to calculate the energy.
In that case, the non-iterative truncation process is not only useful for building a resource-efficient Hamiltonian to simulate dynamics, but it is also beneficial for ground state calculations if the goal is a chemically accurate energy.

\section{Simulation Details} \label{sec:simulation details}
\begin{table*}[t]
\centering
{\renewcommand{\arraystretch}{1.5}%
\small\begin{tabular*}{\linewidth}{l@{\extracolsep{\fill}}*{12}{l}}
\hline
Molecule &  N$_\text{O}$ &  N$_\text{V}$ & Qubits &  \thead{one-body \\ terms} & \thead{two-body \\ terms} & \thead{terms in $\mathbb{I}^0$ }& \thead{terms in $\mathbb{T}^0$} & \thead{$vvvv$ terms \\ variational \\ SQuISH} & \thead{$vvvv$ terms \\non-variational \\ SQuISH} \\ 
\hline
NH$_3$ (cc-pVTZ) &5 &67 & 144 &\num{5.18e3} & \num{2.69e7}  & \num{6.72e6} & \num{2.02e7} & \num{1.93e5} & \num{2.11e6} \\ \grayline
HF (cc-pVQZ) &5 &80 & 170 & \num{7.23e3} & \num{5.22e7} & \num{1.12e7} & \num{4.70e7} & \num{3.28e4} & \num{1.12e6} \\ \grayline
LiH (cc-pVQZ) &2 &83 & 170 &\num{7.23e3} & \num{5.22e7} & \num{4.74e6} & \num{4.75e7} & \num{1.91e4} & \num{2.25e5} \\ \grayline
F$_2$ (cc-pVQZ) &9 &101 & 220 & \num{1.21e4} & \num{1.46e8} & \num{4.23e7} & \num{1.04e8} & \num{1.93e5} & \num{2.17e6} \\ \grayline
C$_2$ (cc-pVQZ) &6 &104 & 220 &\num{1.21e4} & \num{1.46e8} & \num{2.94e7} & \num{1.17e8} & \num{2.05e5} & \num{2.29e6} \\ \grayline
H$_2$ (cc-pV5Z) &1 &109 & 220 &\num{1.21e4} & \num{1.46e8} & \num{5.25e6} & \num{1.41e8} & \num{4.2e3} & \num{2.26e4} \\ \grayline
H$_2$O (cc-pVQZ) &5 &110 & 230 &\num{1.32e4} & \num{1.75e8} & \num{2.85e7} & \num{1.46e8} & \num{3.62e5} & \num{5.70e6} \\ \grayline

BeH$_2$ (cc-pVQZ) &3 &112 & 230 & \num{1.32e4} & \num{1.74e8}  & \num{1.75e7} & \num{1.57e8} & \num{4.04e4} & \num{3.97e5} \\
\hline
\end{tabular*}} \quad
\caption{\label{tab:CCtable} This table shows the data for coupled cluster variational and non-variational SQuISH tests using PyCC. 
The N$_\text{O}$ and N$_\text{V}$ columns show the number of occupied and virtual spatial orbitals, respectively. 
The qubits column shows the number of qubits required to simulate the molecule on a quantum computer. 
The one- and two-body terms columns show how many total terms are in $\hat{H}$. 
The terms in $\mathbb{I}^0$ column shows the number of ERI terms in the initial truncated Hamiltonian, and the terms in $\mathbb{T}^0$ column shows the number of terms in $\hat{H}-\hat{H}_T$.
The variational and non-variational SQuISH columns illustrate the number of $vvvv$ needed to achieve chemical accuracy.  
We use Feller geometries for H$_2$O, HF, NH$_3$ F$_2$, and LiH, and we use NIST calculated geometries at equilibrium for $\ce{C2}$, $\ce{H2}$, and $\ce{BeH2}$ \cite{Feller2008, cccbdb1999}.}
\end{table*}

We generate ground state wavefunctions using configuration interaction for systems on the order of 10 qubits and coupled cluster for systems that would require on the order of 100+ qubits to test our approach.   
These simulations serve as a test bed for the performance we can expect to achieve in practice with regard to the compression of a Hamiltonian.   
In practice, various methods could 
be used to generate the iterative wavefunction during the SQuISH algorithm, such as time evolution, phase estimation, adiabatic state preparation \cite{aspuru-guzik2005, Crosson_2021,Hauke_2020, Brady2021, Kremenetski_2021}, and QAOA~\cite{Kremenetski_2021_2}.

As we consider larger basis sets, the virtual space dominates. 
The majority of the terms in the Hamiltonian come from the two-body virtual terms. 
Thus, we chose the initial Hamiltonian such that it includes all terms except two-body virtual terms in Step 1. 
The initial Hamiltonian, $\hat{H}_T^0$, is defined using the initial set of terms shown below,

\begin{equation}\label{CC_init}
\begin{split}
    &\mathbb{I}^0 = \{(p,q)\} \cup \\
    &\{(p,q,r,s) : (p,q,r,s) \notin \{(v,v,v,v)\}\}.
\end{split}
\end{equation}
\noindent Here, $v$ denotes the virtual orbitals. Similar to the two benchmarks described in the previous section, in Step 2 we have the option of calculating the energy non-variationally  (i.e., Eqn.~\eqref{nv_energy}) or variationally (i.e., Eqn.~\eqref{v_energy}). We implement and compare both non-variational and variational SQuISH.

We use PyCC~\cite{PyCC2021} to run coupled cluster and calculate the energy for the truncated Hamiltonians with our method.  
During this approach, we also generate the density matrices used for the ranking and energy calculations. The ranking equations are based on the reduced density matrices (Eqns.~\ref{1rdm} and \ref{2rdm}).  
The non-variational energy is calculated with only the subset of terms included in the truncated Hamiltonian,
\begin{equation} \label{CC_T_energy_nv}
    E_{CCSD_{T,{\text{nv}}}}^k = \sum_{p,q} h_{pq}\gamma_{p,q} + \sum_{p,q,r,s \in \mathbb{I}^k} h_{pqrs}\Gamma_{p,q,r,s}^k.
\end{equation}
As mentioned above, one can calculate a variational energy, with respect to the full Hamiltonian, even when using the ground state of the truncated Hamiltonian.  
We modify the summation to include all of the elements in $\hat{H}$, not just $\hat{H}_T$ 
\begin{equation} \label{CC_T_energy_v}
    E_{CCSD_{T,{\text{v}}}}^k = \sum_{p,q} h_{pq}\gamma_{p,q} + \sum_{p,q,r,s}   h_{pqrs}\Gamma_{p,q,r,s}^k.
\end{equation}
We note that while in practice, the SQuISH termination condition in Step 3 is based on the error calculated using the energy from the previous iteration (i.e., $\abs{E_T^{k-1} - E_T^k}$), we use the exact energy (i.e., $\abs{E_0 - E_T^k}$) for testing purposes.

In Step 4, we calculate the energetic importance of each term in $\mathbb{T}^k$ using the 2RDMs
\begin{equation} \label{CC_term_expval}
    \epsilon_{i} =  h_{pqrs}\Gamma_{p,q,r,s}^k \text{ for $(p,q,r,s) \in \mathbb{T}^k$}.
\end{equation} 
We rank the terms using Eqn.~\eqref{CC_term_expval} based on the energetic contribution in Step 5.
In Step 6, we start with a small $m$ with respect to system size and, after the first iteration, continually increase it by an order of magnitude every few iterations.

For small systems, we test SQuISH using Qiskit~\cite{Qiskit} with configuration interaction, which is an exact method (unlike coupled cluster).
Due to current limitations, we use small systems to do the configuration interaction truncation.
The molecules we use to perform our calculations require 12 qubits. 
As a result, the truncation space is relatively small.

\section{Results and Discussion} \label{sec:results}

\subsection{Coupled Cluster Results} \label{CCResults}

\begin{table*}[t]\label{FCItable}
\centering
\small\begin{tabular*}{\linewidth}{l@{\extracolsep{\fill}}*{12}{l}}
\hline
Molecule &  N$_\text{O}$ &  N$_\text{V}$  & \thead{one-body \\ terms} & \thead{two-body \\ terms} & \thead{terms in $\mathbb{I}^0$}& \thead{terms in $\mathbb{T}^0$} & \thead{$vvvv$ terms \\ variational \\ SQuISH} & \thead{
$vvvv$ terms \\non-variational \\ SQuISH} & \thead{$\abs{\bra{\Psi_{T,0}^k}\ket{\Psi_0}}$ } \\

\hline
\ce{H2} (cc-pVDZ) &1 &5 & 36 &1,296 &   671 & 625 & 6 & 55 & 0.99951506 \\ \grayline
\ce{H3+} (cc-pVDZ) &1 &5 & 36 &1,296 & 671 & 625 & 16 & 33 & 0.99974916 \\ \grayline
\ce{LiH} (STO-3G) &2 &4 & 36 &1,296 &  1040 & 256 & 10 & 10 & 0.99990412 \\
\hline
\end{tabular*}
\caption{\label{tab:FCItable} 
This table provides data from the variational and non-variational SQuISH tests using Qiskit. 
The N$_\text{O}$ and N$_\text{V}$ columns show the number of occupied and virtual spatial orbitals, respectively. 
The one- and two-body terms columns show how many total terms are in $\hat{H}$. 
The $\mathbb{I}^0$ column shows the number of ERI terms in the initial truncated Hamiltonian, and the $\mathbb{T}^0$ column shows the number of terms in $\hat{H}-\hat{H}_T$. 
The variational and non-variational SQuISH columns show how many $vvvv$ ERI terms are needed to achieve chemical accuracy after applying SQuISH. 
The last column shows the overlap between the accurate ground state wavefunction and the ground state wavefunction of the variational SQuISH truncated Hamiltonian at chemical accuracy. For each molecule, we use the equilibrium calculated geometries from NIST \cite{cccbdb1999}.}
\end{table*}

We use variational and non-variational SQuISH with coupled cluster states to truncate the Hamiltonians of the following systems: HF, LiH, F$_2$,  C$_2$, H$_2$O and BeH$_2$ in the cc-pVQZ basis, NH$_3$ in the cc-pPVTZ basis, and H$_2$ in the cc-pV5Z basis. 
For each molecule, we evaluate the number of two-body virtual terms that need to be included in the truncated Hamiltonian to reach chemical accuracy after applying SQuISH. 
Table \ref{tab:CCtable} contains the data illustrating the total number of terms in each Hamiltonian and the number of terms after applying the SQuISH algorithm. 
The number of terms is reduced by between two to five orders of magnitude after applying SQuISH.
We also see that calculating the energy variationally improves the truncation by at least an order of magnitude.

SQuISH not only reduces the number of Hamiltonian terms but can also have a  favorable effect on the Hamiltonian norm,  with different norms yielding resource estimates for different algorithms~\cite{childs2021theory}. 
Such norms commonly arising in matrix analysis can often be bounded in terms of each other.
Here we consider the sum of the absolute values of the Hamiltonian coefficients as an indicative quantity, i.e., the 1-norm of the vector of coefficients, which we denote as

\begin{equation} \label{one norm}
    \vertiiii{\hat{H}}_1 := \sum_{p,q} \abs{h_{pq}} + \frac{1}{2} \sum_{p,q,r,s} \abs{h_{pqrs}}.
\end{equation}
This quantity is used in the calculations shown in Table \ref{tab:norms}.
We calculate and compare it for the non-truncated Hamiltonians, the chemically accurate SQuISH truncated Hamiltonians, and the coefficient ranking truncated Hamiltonians.

\begin{table}[H]
\begin{center}
{\renewcommand{\arraystretch}{1.5}
\setlength{\tabcolsep}{0.5em}
\begin{tabular}{ l l l l }
\hline
Molecule & ${\vertiiii{\hat{H}}}_1$ & \thead{SQuISH \\ ${\vertiiii{\hat{H}_T}}_1$} & \thead{Coefficient  Ranking \\ ${\vertiiii{\hat{H}_T}}_1$}\\ \hline
 \ce{NH3} (cc-pVTZ) & \num{3.26e4} & \num{1.18e4} & \num{2.18e4}\\ 
 \grayline
 \ce{HF} (cc-pVQZ) & \num{4.57e4}& \num{1.81e4} & \num{2.35e4}\\
 \grayline
 \ce{LiH} (cc-pVQZ) & \num{1.65e4} & \num{2.15e3} & \num{7.40e3} \\
 \grayline
 \ce{F2} (cc-pVQZ) & \num{7.91e4} & \num{3.36e4} & \num{5.54e4}\\
 \grayline
\ce{C2} (cc-pVQZ) & \num{5.66e4} & \num{1.92e4} & \num{3.90e4} \\
 \grayline
 \ce{H2} (cc-pV5Z) & \num{7.25e4} & \num{4.28e3} & \num{2.10e4} \\
 \grayline
 \ce{H2O} (cc-pVQZ) & \num{1.01e5} & \num{2.71e4} & \num{5.45e4} \\
 \grayline
 \ce{BeH2} (cc-pVQZ) & \num{4.88e4} & \num{8.16e3} & \num{2.71e4} \\
 \hline
\end{tabular}}
\end{center}
\caption{\label{tab:norms} This table shows the one norms of the full and truncated Hamiltonians of each molecule tested. $\vertiiii{\hat{H}_T}_1$ is calculated for both SQuISH and the coefficient ranking scheme truncation once chemical accuracy is reached.}
\end{table}

\begin{figure}[H]
    \centering
    \begin{subfigure}[b]{\linewidth}
        \includegraphics[width=\textwidth]{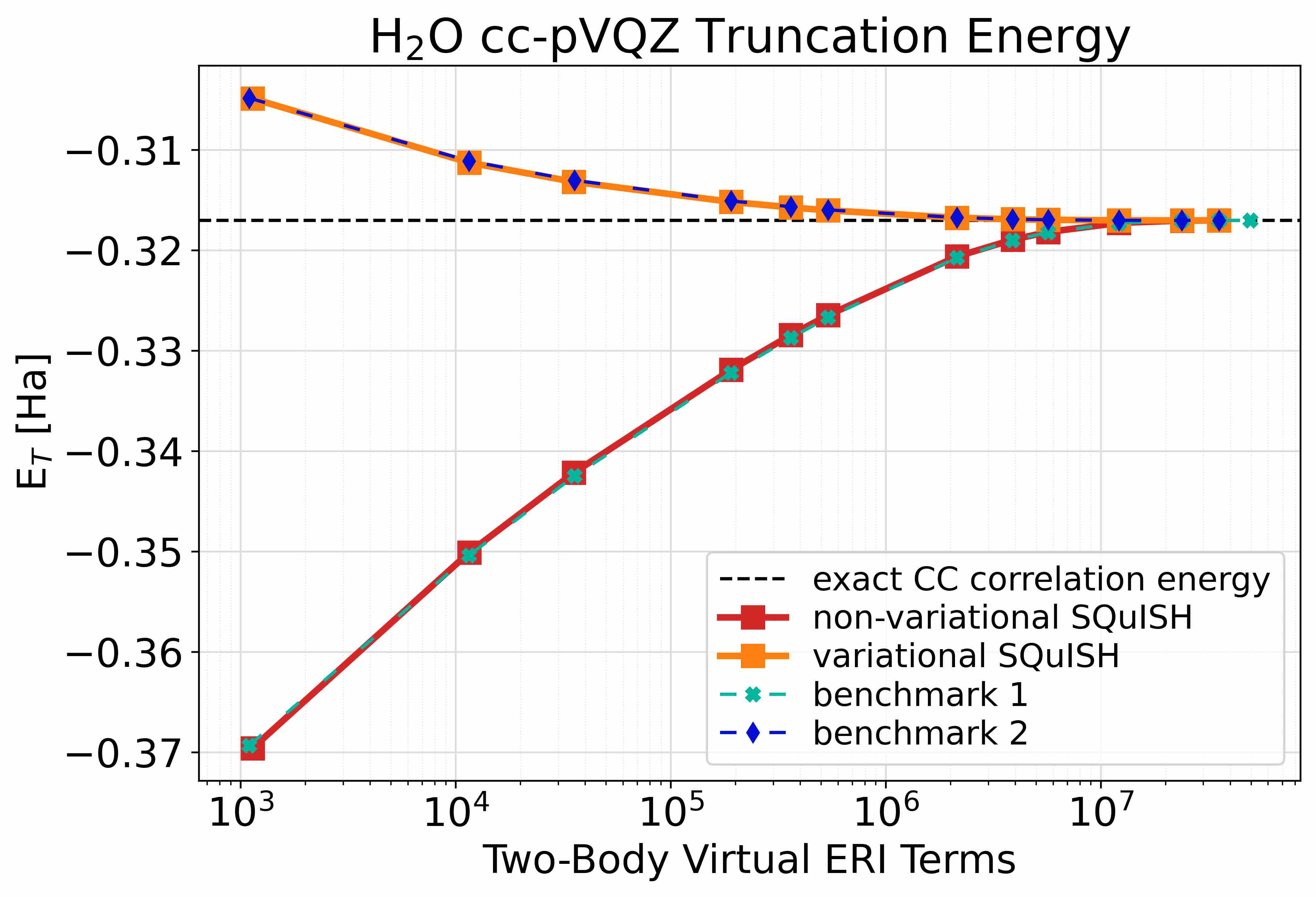}
        \caption{Semi-logarithmic energy plot}
        \label{fig:cc_H2O_energy}
    \end{subfigure}
    \hfill
    \centering    
    \begin{subfigure}[b]{\linewidth}
        \includegraphics[width=\textwidth]{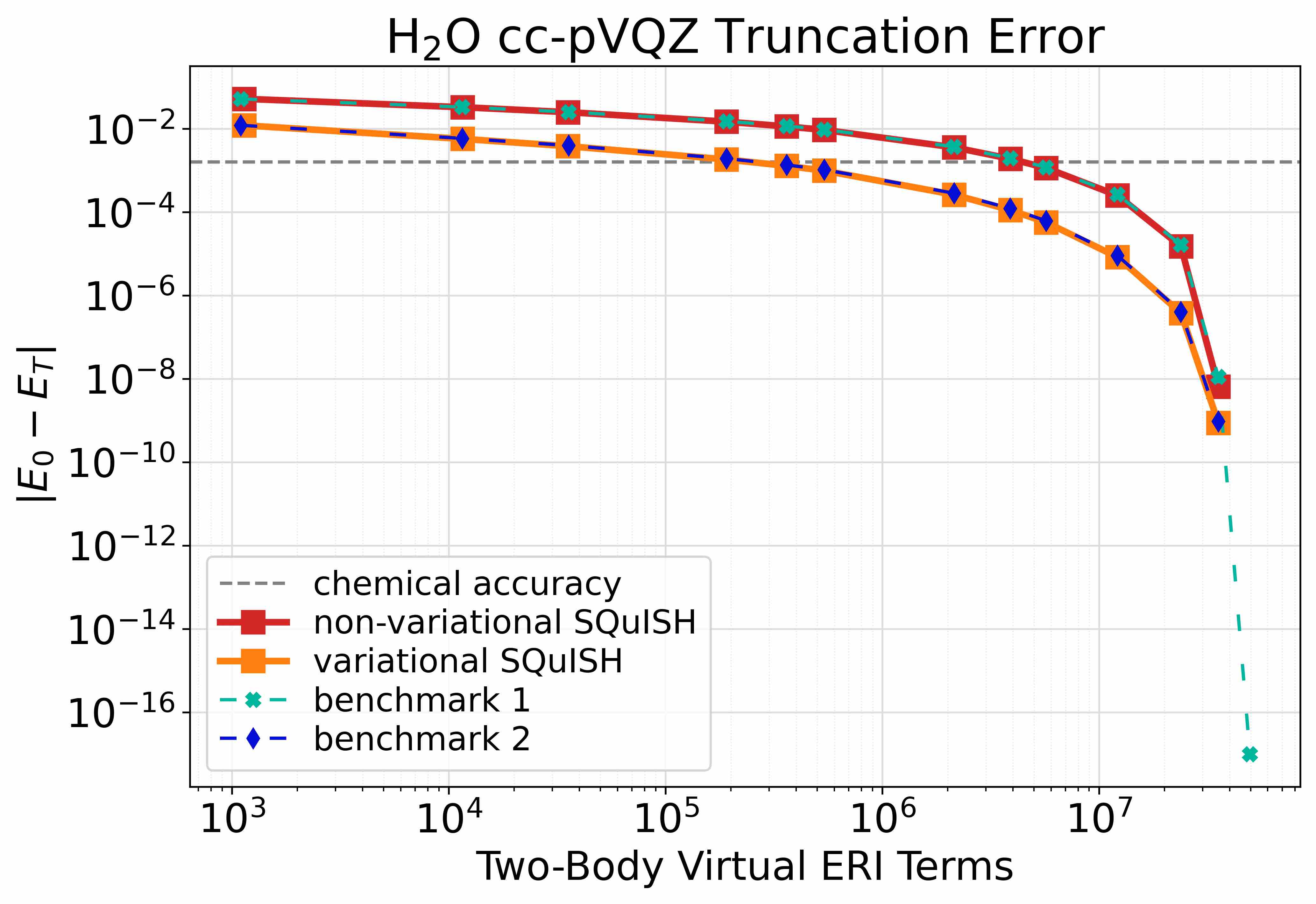}
        \caption{Semi-logarithmic error plot}
        \label{fig:cc_H2O_error}
    \end{subfigure}
    \caption{Coupled-cluster \ce{H2O} in the cc-pVQZ basis. 
    The energy on the plots is the coupled cluster correlation energy. 
    Each point on the plots represents one self-consistent iteration. 
    SQuISH and the benchmarks terminate once the approximate energy is within \num{1e-8} Hartrees of the exact energy.}
    \label{cc_h2o}
\end{figure}

Fig.~\ref{cc_h2o} presents more detailed results at each iteration using coupled cluster calculations for H$_2$O in the cc-pVQZ basis.
We use the energy and error plots to evaluate the two benchmark comparisons, variational SQuISH, and non-variational SQuISH using the $\ce{H2O}$ data.  Considering Fig.~\ref{fig:cc_H2O_energy}, we see that Benchmark 1 and Benchmark 2 almost completely overlap variational SQuISH and non-variational SQuISH, respectively. 
This implies that using an approximate wavefunction for the ranking is nearly as good as using the exact wavefunction. 

The high accuracy we get from our approximate Hamiltonians and approximate wavefunctions is not surprising given that classical methods have shown similar convergence properties in slightly different contexts, such as the ASCI algorithm~\cite{Tubman2016}. 
Given SQuISH behaves so similarly to the benchmarks, we conclude that $\ket{\Psi_{T,0}^k}$ contains enough information at each step to find the next set of important terms to add in the Hamiltonian.

The results exemplify the advantage in calculating the energy variationally. 
However, there is a trade-off between a faster convergence and the number of measurements. 
If the energy is calculated non-variationally and we use previously discussed techniques, such as shadow tomography, in Step 4, we can reduce the number of measurements we need to make on a quantum computer in exchange for accuracy.

\subsection{Configuration Interaction Results} \label{fci_results}

We test variational and non-variational SQuISH for \ce{H2}, \ce{H3+}, \ce{LiH} for various basis sets using Qiskit. 
Table \ref{tab:FCItable} illustrates how many terms are required to achieve chemical accuracy for the molecules tested. 

We use full configuration interaction for \ce{LiH} in the STO-3G basis and truncate \ce{H2} and \ce{H3+} to 6 spacial orbitals in the cc-pVDZ basis. 
Although the number of ERIs are fairly small, we still substantially truncate the number of terms after applying SQuISH. 
The variational truncation slightly outperforms the non-variational truncation when applied to \ce{H2} and \ce{H3+}, and they perform about the same when applied to \ce{LiH}. 
The advantage of variational SQuISH is reduced here in comparison to the coupled cluster results because we are using very small systems. 
We include results showing the wavefunction overlap once the ground state energy is within chemical accuracy. 
Additionally, we test the non-iterative truncation with VQE using a UCC with singles and doubles (UCCSD) ansatz on the \ce{H2} Hamiltonian.

\begin{figure}[H]
    \centering
    \begin{subfigure}[b]{\linewidth}
        \includegraphics[width=\textwidth]{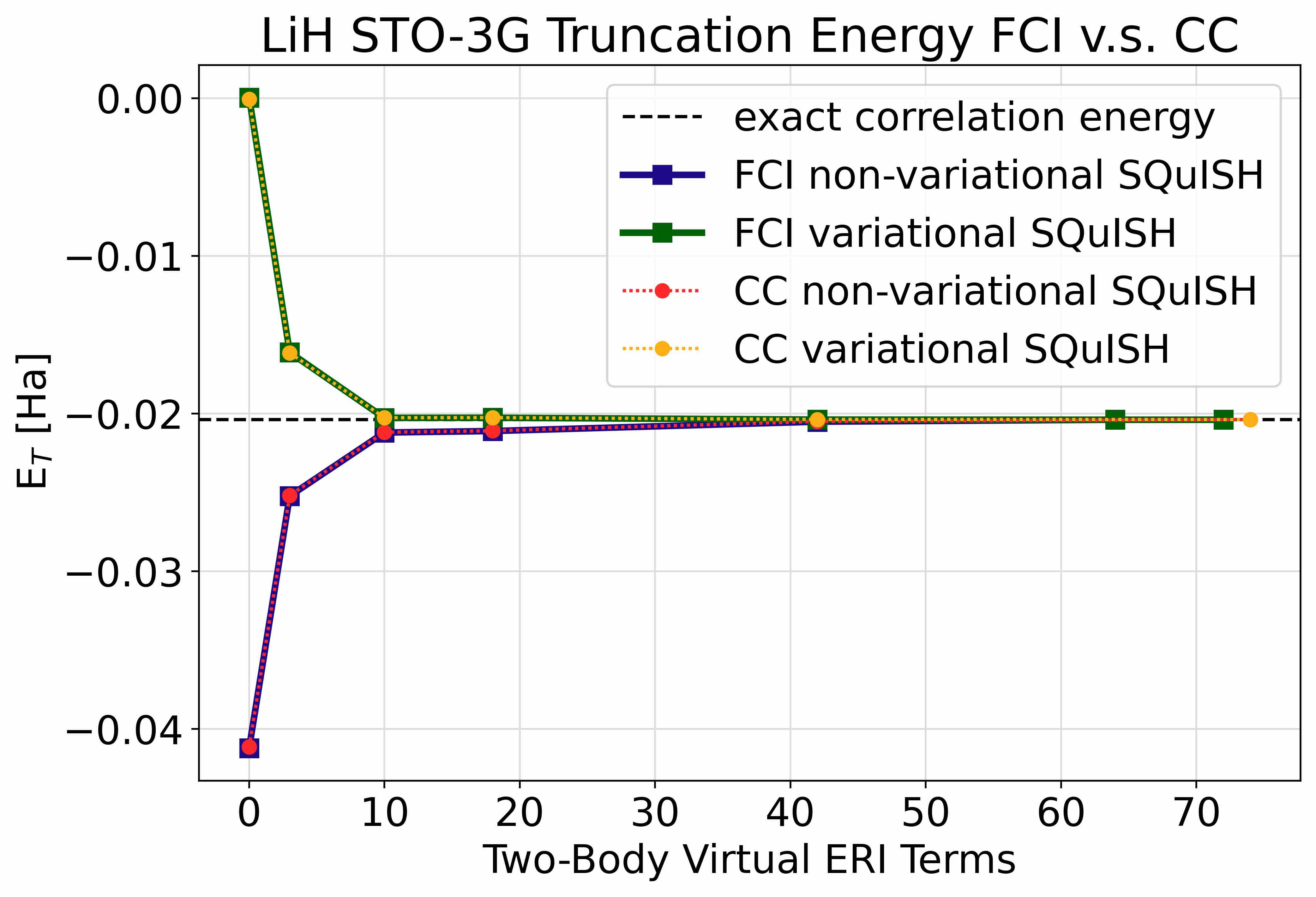}
        \caption{Energy plot}
        \label{fig:lih_energy_comparison}
    \end{subfigure}
    \hfill
    \centering    
    \begin{subfigure}[b]{\linewidth}
        \includegraphics[width=\textwidth]{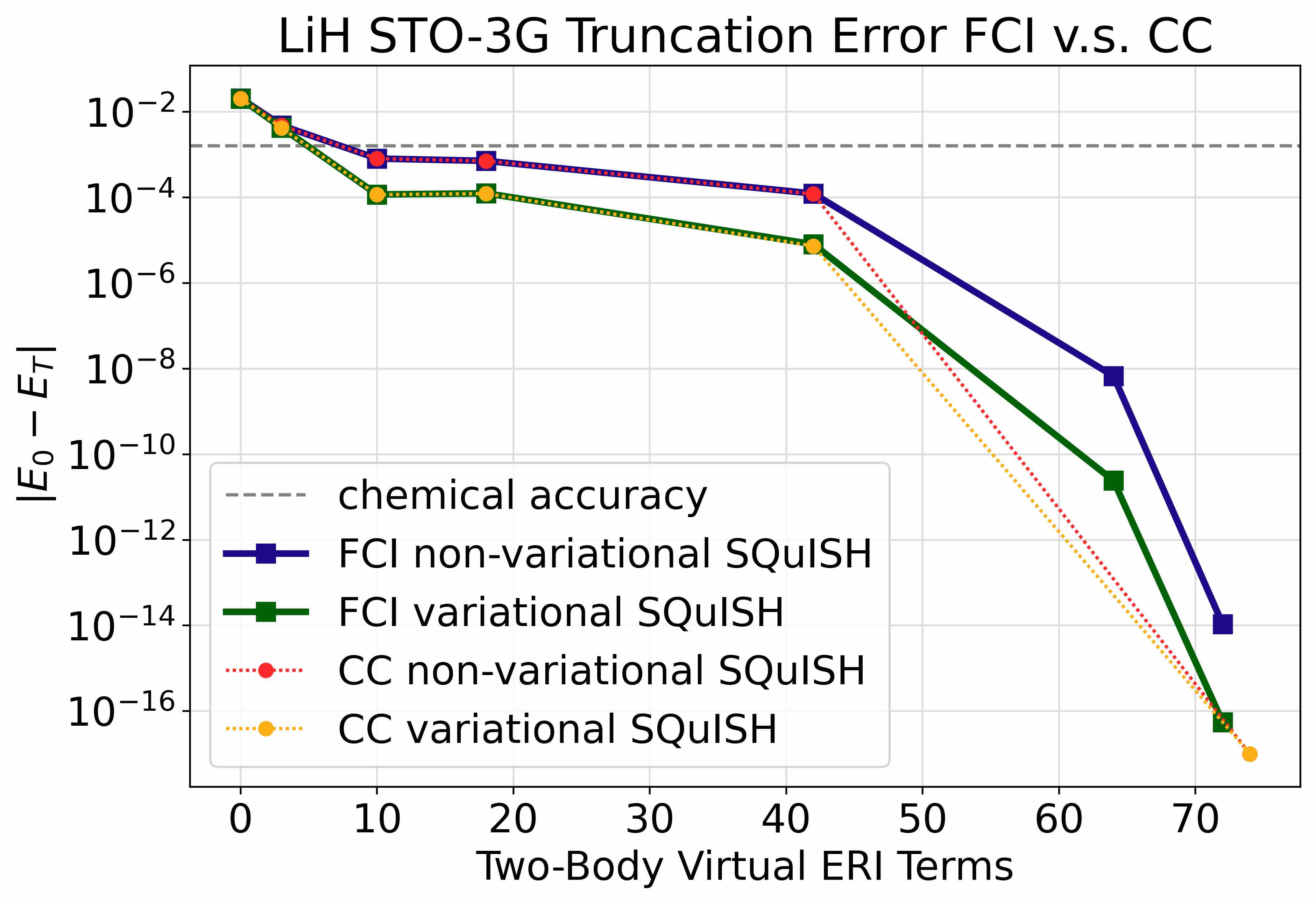}
        \caption{Error plot}
        \label{fig:lih_error_comparison}
    \end{subfigure}
    \caption{A comparison of the full configuration interaction and coupled cluster tests of SQuISH using \ce{LiH} in the STO-3g basis.
    The dark green and light orange compare the variational results.
    The dark blue and light red lines compare the non-variational results.
    We plot the correlation energy.
    SQuISH and the benchmarks terminate once the approximate energy is within \num{1e-12} Ha of the exact energy.}
    \label{lih_cc_fci}
\end{figure}

As a direct comparison, we test SQuISH using $\ce{LiH}$ in the STO-3G basis for coupled cluster and full configuration interaction.
A main goal here is to show that the approximations from coupled cluster do not significantly change from what can be expected from a quantum computer, which is closely represented by the configuration interaction results. 
Our results, in Fig.~\ref{lih_cc_fci},  do indeed look very similar between the two approaches, which gives us confidence that our coupled cluster results (in Section \ref{CCResults}) for larger systems reasonably demonstrate how well the SQuISH compression technique works on larger Hamiltonians. For comparisons between accurate configuration interaction and coupled cluster simulations, see~\cite{Tubman2018}. 

\begin{figure}[H]
    \centering
    \begin{subfigure}[b]{\linewidth}
        \includegraphics[width=\textwidth]{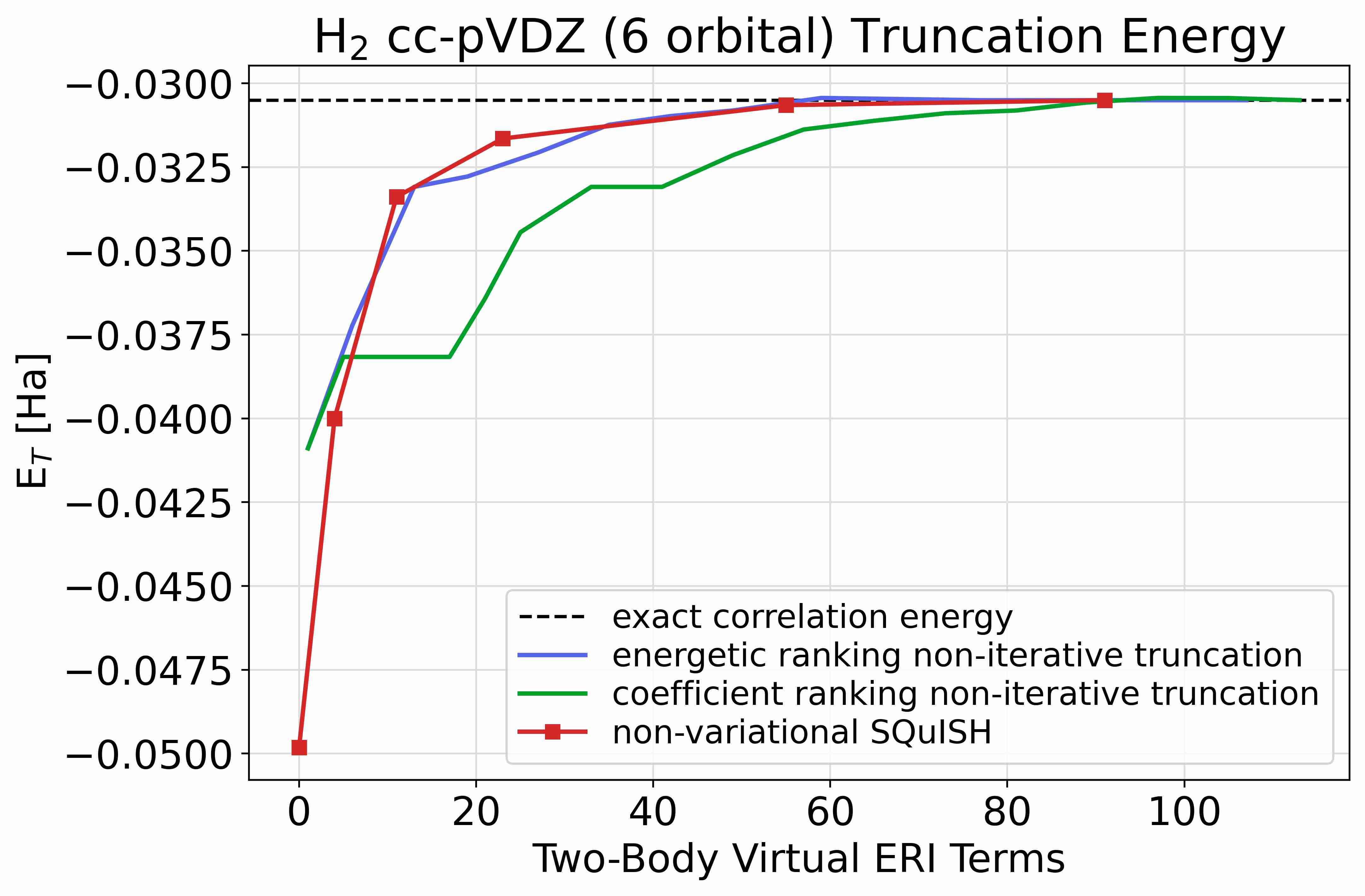}
        \caption{Energy plot}
        \label{fig:non-iterative energy}
    \end{subfigure}
    \hfill
    \centering    
    \begin{subfigure}[b]{\linewidth}
        \includegraphics[width=\textwidth]{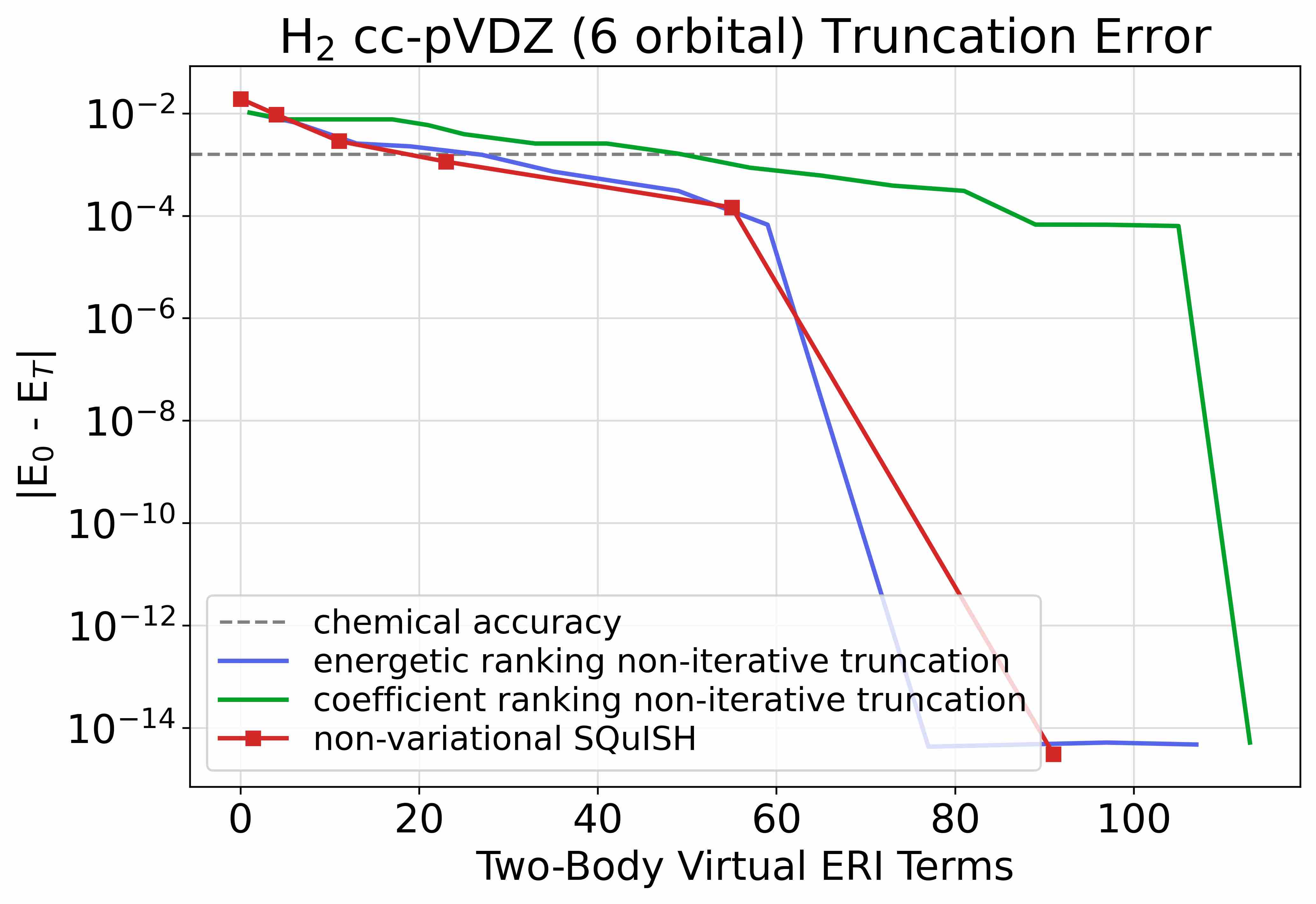}
        \caption{Error plot}
        \label{fig:non-iterative error}
    \end{subfigure}
    \caption{A comparison of non-iterative truncation using VQE and non-variational SQuISH using 6 orbital \ce{H2} in the cc-pVDZ basis.
    The purple line shows the use of the energetic ranking scheme with an optimized UCCSD ansatz.
    The green line shows the coefficient ranking scheme.
    The red line shows non-variational SQuISH, where each point represents one self-consistent iteration. }
    \label{Non-iterative Truncation}
\end{figure}

The results of the non-iterative truncation are shown in Fig.~\ref{Non-iterative Truncation}. 
We test it on the \ce{H2} Hamiltonian by first optimizing a parameterized UCCSD ansatz with VQE to an accuracy of \num{2.02e-2}, which is above chemical accuracy. 
Then we rank terms based on energetic contributions and build the truncated Hamiltonian.
We also compare against the coefficient ranking, in which we determine importance based on $\abs{h_{pqrs}}$. 
For both of these methods, we use the exact ground state wavefunction to calculate the energy at the end (i.e., $\expval{\hat{H}_T}{\Psi_0})$. 
We also add non-variational SQuISH to the plot as a comparison metric. 

The non-iterative truncation with the energetic ranking performs fairly similarly to SQuISH.
As we expected, the VQE trial wavefunction performed well when calculating the energetic contributions even though we terminated VQE prior to achieving a chemically accurate solution.
Moreover, the fact that both SQuISH and the energetic ranking non-iterative truncation outperform the coefficient ranking truncation further validates the argument for an energetic ranking scheme.

\subsection{Energetic versus Coefficient Ranking}
Here we will provide data to support the claim that ranking Hamiltonian terms based on energetic contributions is better than ranking based on the value of the Hamiltonian coefficients alone. 
We use coupled cluster states with Benchmark 1 for the comparison.

\begin{figure}[H]
    \centering
    \begin{subfigure}[b]{\linewidth}
        \includegraphics[width=\textwidth]{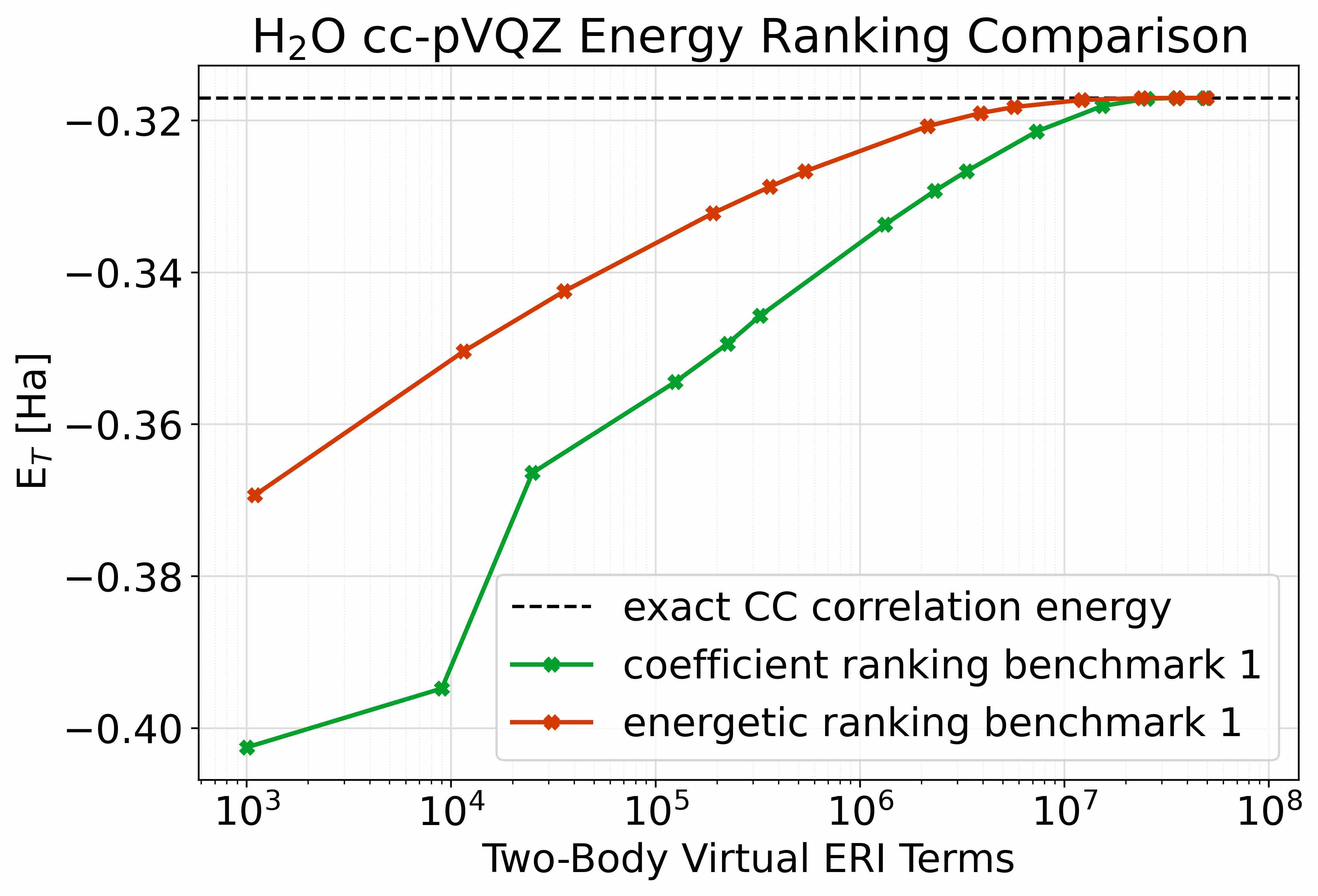}
        \caption{A semi-logarithmic plot of the number of two-body virtual terms included in the truncated Hamiltonian at each iteration versus the approximate coupled cluster correlation energy.}
        \label{fig:ranking_plot energy}
    \end{subfigure}
    \hfill
    \centering    
    \begin{subfigure}[b]{\linewidth}
        \includegraphics[width=\textwidth]{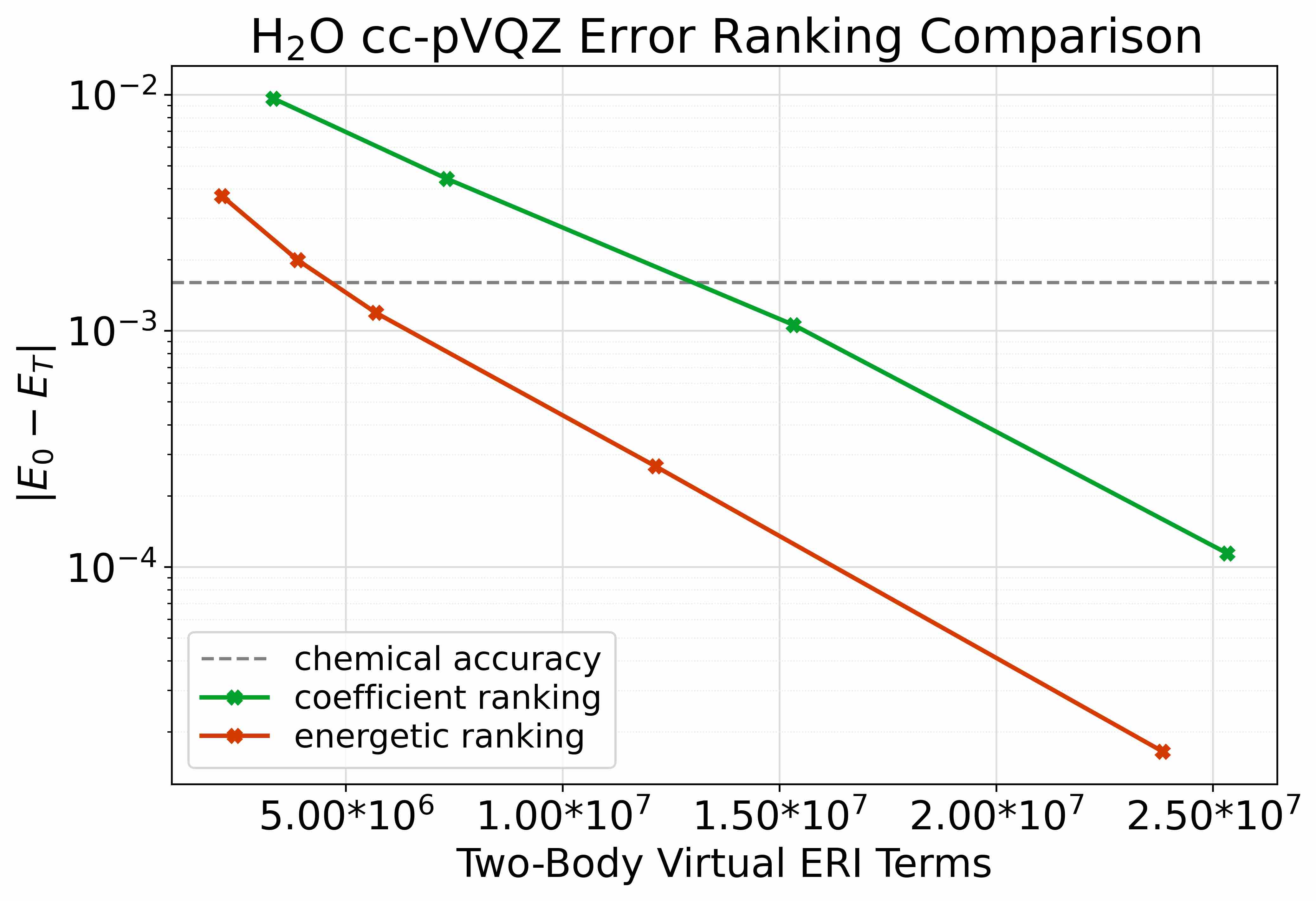}
        \caption{Semi logarithmic plot of the number of two-body virtual terms versus the absolute value of approximate energy error near chemical accuracy.}
        \label{fig:ranking_plot_error}
    \end{subfigure}
    \caption{Energetic ranking scheme versus coefficient ranking scheme using coupled cluster Benchmark 1 with H$_2$O in the cc-pVQZ basis. The algorithm terminates once the energy is within $\num{1e-8}$ of the accurate energy.}
    \label{ranking_plot_comparison}
\end{figure}

For each approach, we initialize the truncated Hamiltonian with Eqn.~\eqref{CC_init}, calculate the energies variationally, and check for convergence in Step 3. The difference between these tests  occurs in Steps 4 and 5. 
For the coefficient ranking, we have nothing to calculate in Step 4, as we have all the coefficients for the Hamiltonian already calculated. 
In Step 5 we rank the terms, $h_{pqrs}\hat{a}_p^\dagger \hat{a}_q^\dagger \hat{a}_r \hat{a}_s$, based on absolute value of the coefficients, $\abs{h_{pqrs}}$.

We test the two ranking methods here using H$_2$O in the cc-pVQZ basis.
The significant advantage we gain by using the expectation value ranking is evident in Fig.~\ref{ranking_plot_comparison}.
Fig.~\ref{fig:ranking_plot_error} shows the few SQuISH iterations before and after the energy converges to chemical accuracy.
One can see that the energetic ranking requires an order of magnitude fewer terms compared to coefficient ranking to reach a chemically accurate solution.
For larger systems, we expect our approach to lead to even more substantial reductions. 

\section{Conclusion and Outlook} \label{sec:conclusion}
The field of quantum computing has seen an extensive number of proposed approaches to reduce the Hamiltonian complexity with the intention of making time evolution and related approaches feasible using fewer quantum resources.   
While the goal of this work is the same, we take a new approach to accomplish this task that makes use of information from an iteratively improved approximate wavefunction, something that takes advantage of the unique capabilities of quantum computers and has not been used in previous algorithms to our knowledge.   
Using the SQuISH algorithm, we show that this information is, in fact, a key ingredient to creating a compression algorithm, and we demonstrate this on a series of chemistry Hamiltonians.  
The idea that different eigenstates will be sensitive to different terms in the Hamiltonian is evident, and the key proposal in our algorithm is that an approximate trial wavefunction can be good enough to provide improved convergence over canonical approaches. 

Our proposal leads to several expected uses.  
First, we note that although we are hopeful that our compression scheme will be especially advantageous in the NISQ era, quantum advantage, 
it will also be useful in the long term on fault-tolerant hardware.  
The idea that we will run arbitrarily large circuits on fault-tolerant hardware is incorrect, as algorithms can still take days/weeks/months to run, even if they have polynomial scaling.   
Thus compressing a Hamiltonian, which can grow to many gigabytes of memory on classical hardware, will undoubtedly be beneficial in the near term and fault-tolerant era. 
We also want to emphasize that our compression scheme is unique and complementary to a large number of other compression techniques.  
By using the information in the wavefunction, one can pick out better terms to include in the Hamiltonian for other compression schemes as well. 
Thus our method complements other approaches, and we expect this will push us closer to finding a path to quantum advantage for quantum chemistry.

There are several directions for future work to either improve SQuISH or integrate it into various quantum algorithms, including potential applications beyond chemistry. 
To further reduce the complexity of a Hamiltonian, one can explore the best terms to include in the initial Hamiltonian. 
Additionally, by 
refining the algorithm to incorporate state-of-the-art approaches to expectation value estimation, we can 
further reduce the number of measurements the quantum computer needs to make. As mentioned, our approach may be extended to excited states and subspaces. 
Although this paper focused on the ground state problem for quantum chemistry, we expect SQuISH will be useful for other purposes, such as determining ground and excited states for condensed matter systems or any other eigenvalue estimation problem, as well as dynamics.

\section{Acknowledgements}
  We are grateful for support from NASA Ames Research Center.   We acknowledge funding from the NASA ARMD Transformational Tools and Technology (TTT) Project.  Part of this work is funded by U.S. Department of Energy, Office of Science,
National Quantum Information Science Research Centers,
Co-Design Center for Quantum Advantage under Contract No. DE-SC0012704.
  Calculations were performed as part of the XSEDE computational Project No. TG-MCA93S030 on  Bridges-2 at the Pittsburgh supercomputer center. D.C. and S.H. were supported by NASA Academic Mission Services, Contract No. NNA16BD14C.  D.C. participated in the Feynman
Quantum Academy internship program.

\appendix \label{appendix}
\section{Adaptive Sampling Configuration Interaction} \label{sec:ASCI}
The iterative truncation based on energetic contributions used in SQuISH was largely motivated by the Adaptive Sampling Configuration Interaction (ASCI) algorithm. 
We provide a brief overview of ASCI in this section. 
In general, selective configuration interaction procedures find the most important determinants in a Hilbert space and truncate the space such that it only includes those determinants. 
ASCI, in particular, uses a ranking equation reliant on norm coefficients to iteratively update the wavefunction with the most important determinants.

ASCI is motivated by the full configuration interaction quantum Monte Carlo technique \cite{Booth2009}, which was initially introduced as a projector method in imaginary time. 
Let's start by considering a wavefunction, $\ket{\Psi(\tau)}$, where $\tau$ is imaginary time. 
We can expand $\ket{\Psi(\tau)}$ into a linear combination of Slater determinants $\{\ket{D_i}\}$, as follows
\begin{equation}
    \ket{\Psi(\tau)} = \sum_i C_i(\tau)\ket{D_i},
\end{equation}
where $C_i$ are the amplitudes associated with each determinant.
The propagator in imaginary time is
\begin{equation}
    -\frac{d C_i}{d\tau} = (\expval{H}{D_i}) - E_0)C_i + \sum_{i \neq j} \bra{D_i}H\ket{D_j} C_j,
\end{equation}
where $E_0$ is the ground state energy. Setting $\frac{d C_i}{d\tau} = 0$ gives the asymptotic solution of the stationary state for the amplitudes. 
Then, Eqn.~\eqref{eqn3} gives the solutions for each coefficient,
\begin{equation}\label{eqn3}
    C_i = -\frac{\sum_{i \neq j} \bra{D_i}H\ket{D_j} C_j}{\expval{H}{D_i} - E_0}.
\end{equation}

 Using Eqn.~\eqref{eqn3}, ASCI iteratively calculates the importance of the determinants using the wavefunction from the previous iteration, as follows: at iteration $k$ we label the initial input coefficients as $\{C_j^k\}$ and the output coefficients as $\{C_i^{k+1}\}$,
\begin{equation}\label{eqn4}
    C_i^{k+1} = -\frac{\sum_{i \neq j} \bra{D_i}H\ket{D_j} C_j^k}{\expval{H}{D_i} - E_0^k}.
\end{equation}
  Then, the output coefficients, $\{C_i^{k+1}\}$, with the largest magnitudes will be used to construct the larger determinant space, called the target space.
 
At the first iteration, the input wavefunction is set to the Hartree-Fock wavefunction, $C_0^0 = 1$ and $\{C_{j>0}^0 = 0\}$. For $k>0$ the input coefficients are the coefficients found in the previous step, $\{C_j^k\} = \{C_i^{k-1}\}$. Once we have the coefficients, the next step is to build a Hamiltonian in the target space and diagonalize it. 
The procedure is complete if the smallest eigenvalue, $E_0^{k+1}$, converges to the ground state energy within a desired tolerance.
 
SQuISH is inspired by ASCI in that we iteratively improve upon the wavefunction by truncation using a ranking scheme reliant on the Hamiltonian and wavefunction. 
However, rather than using the exact Hamiltonian to calculate an approximate wavefunction, we flip the algorithm and use an approximate Hamiltonian and calculate an exact wavefunction.

\section{Additional Results} \label{sec:more plots}

\begin{figure}[H]
    \includegraphics[width=.24\textwidth]{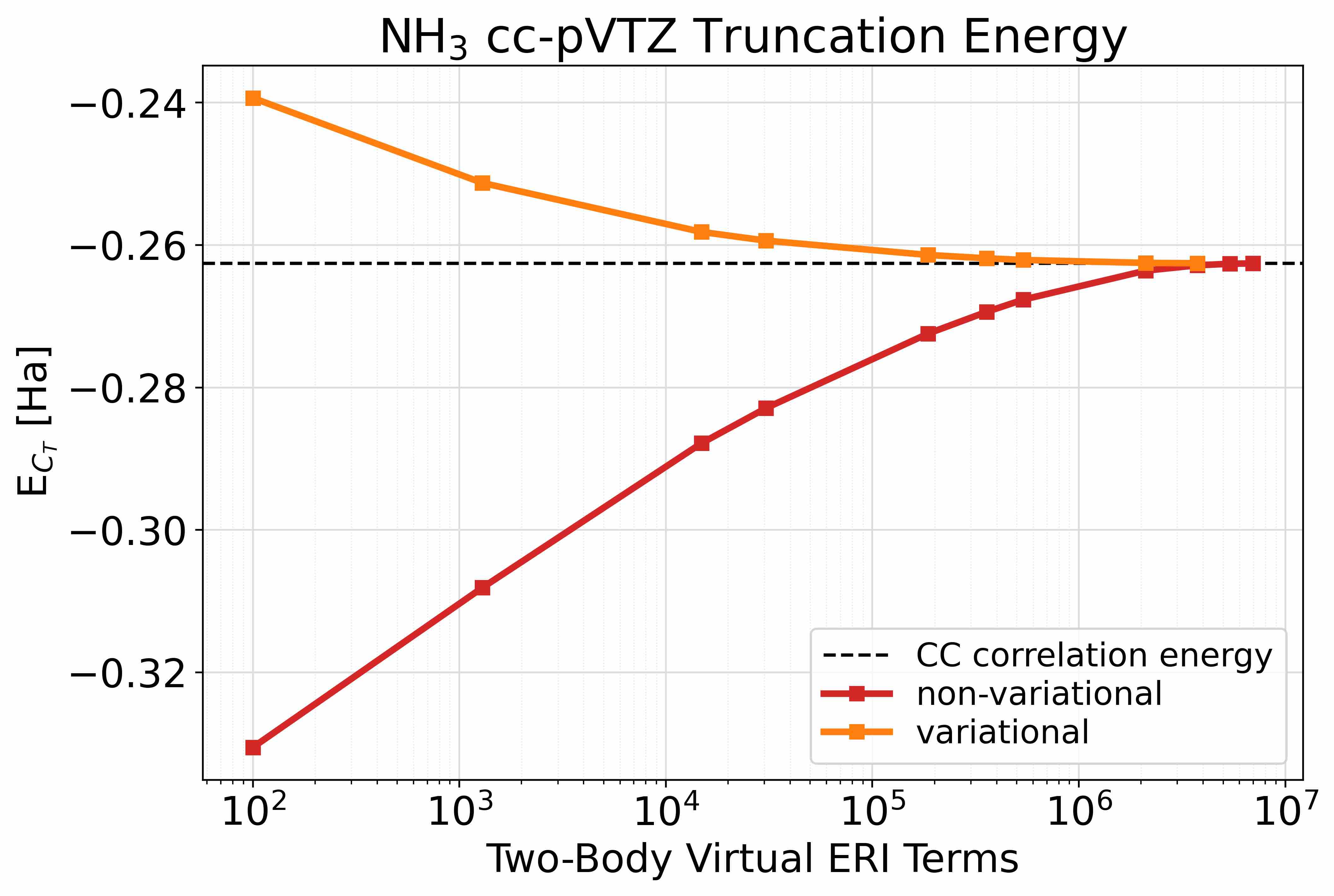}\hfill
    \includegraphics[width=.24\textwidth]{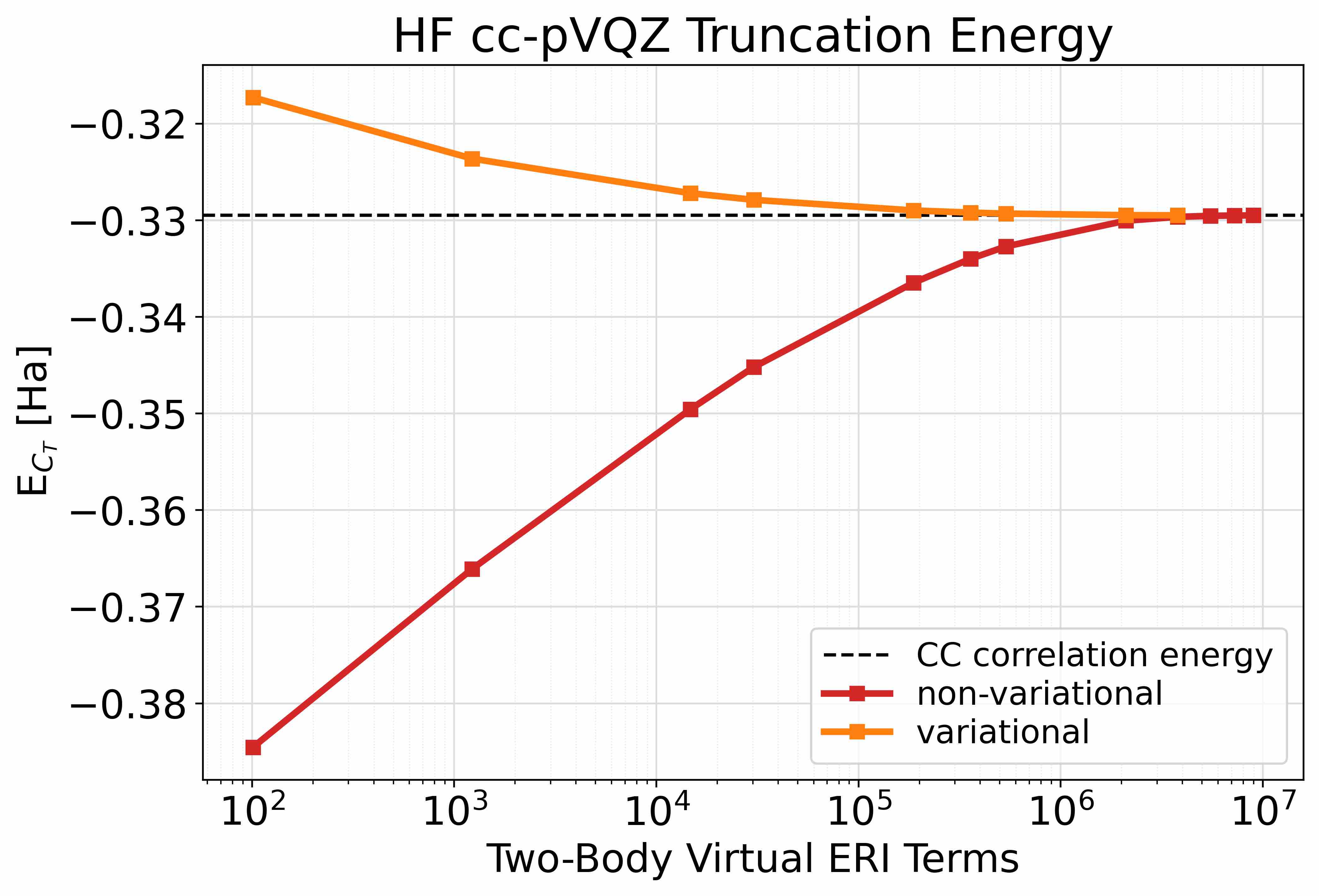}\hfill
    \includegraphics[width=.24\textwidth]{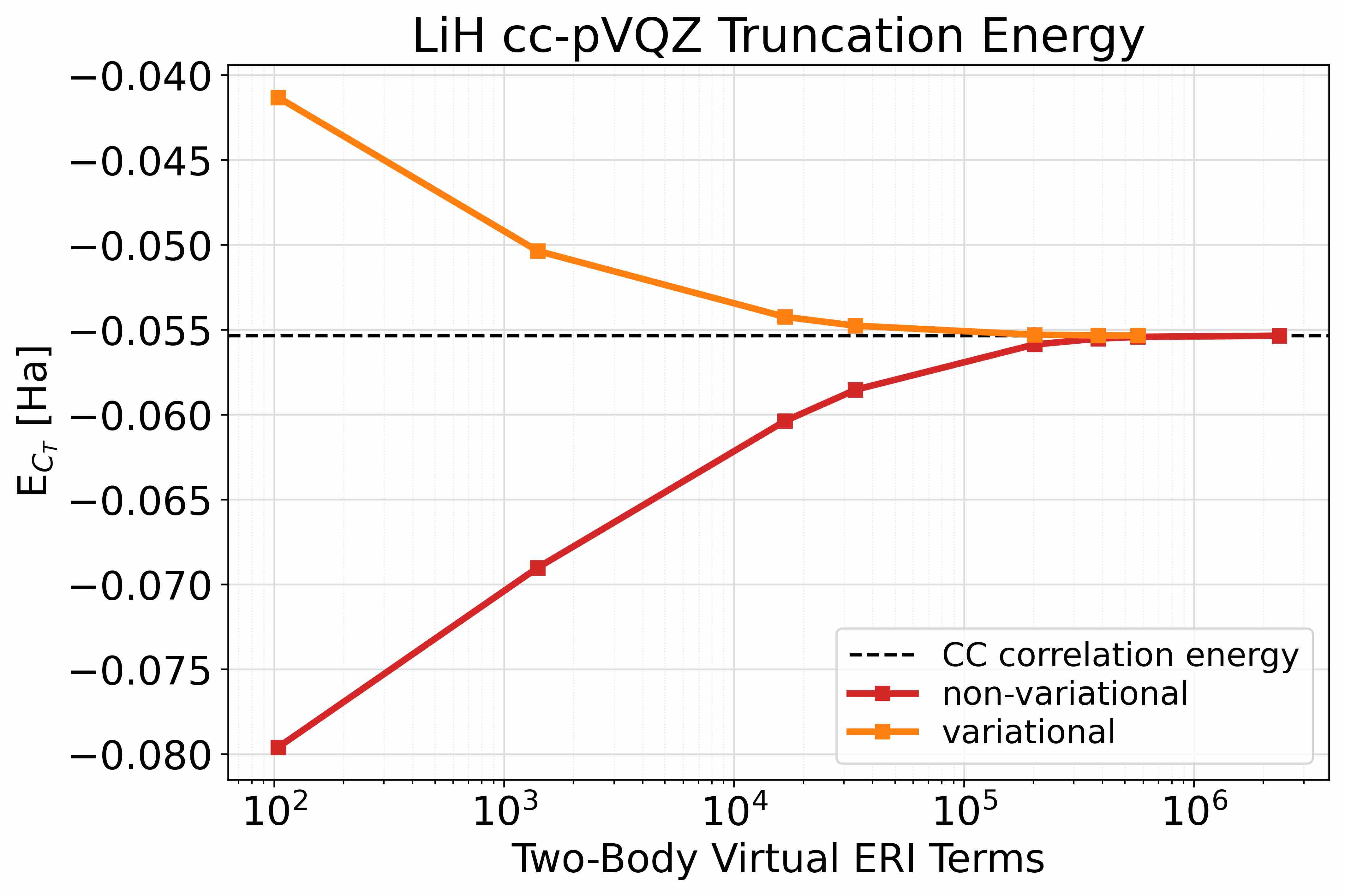}\hfill
    \includegraphics[width=.24\textwidth]{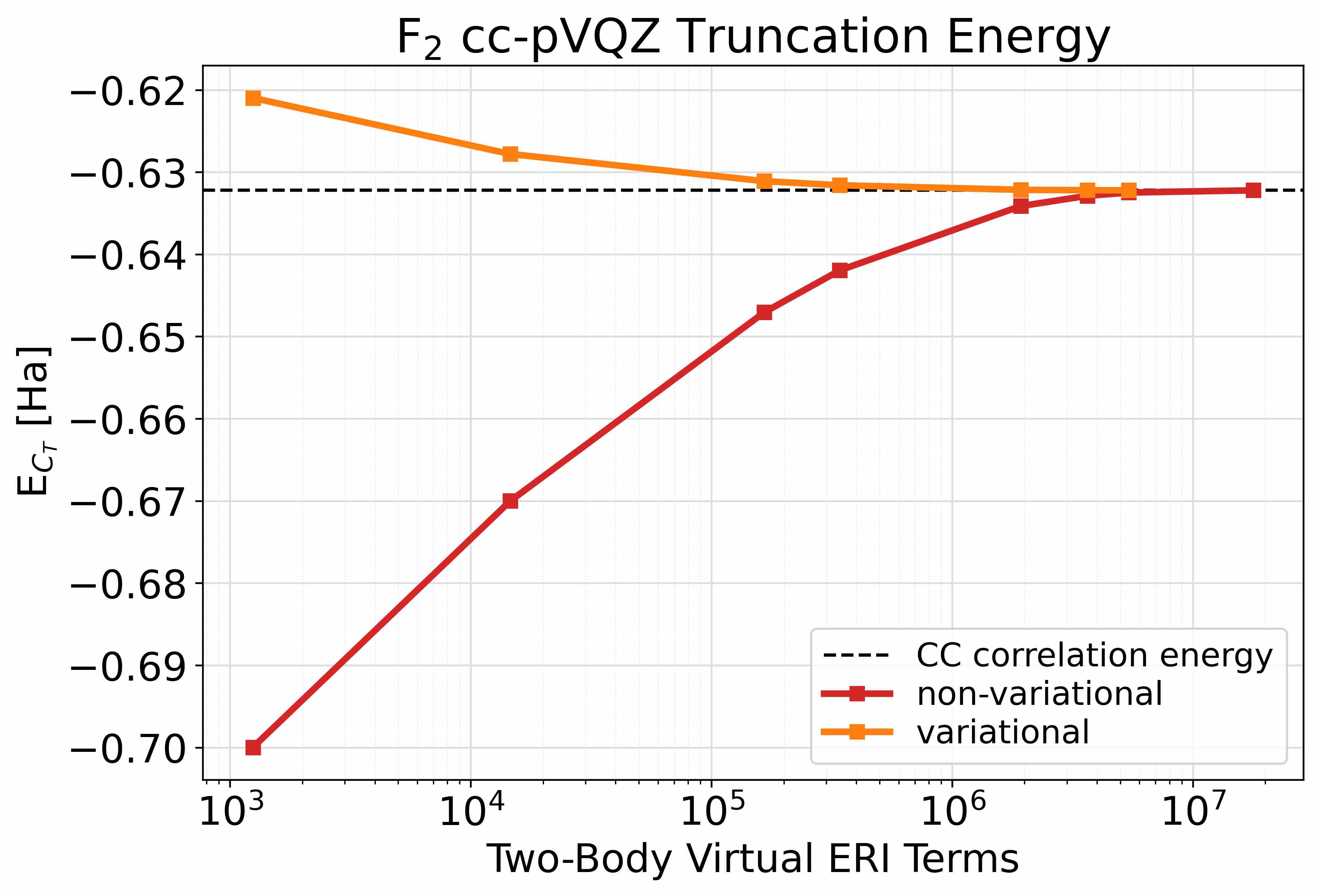}
    \\[\smallskipamount]
    \includegraphics[width=.24\textwidth]{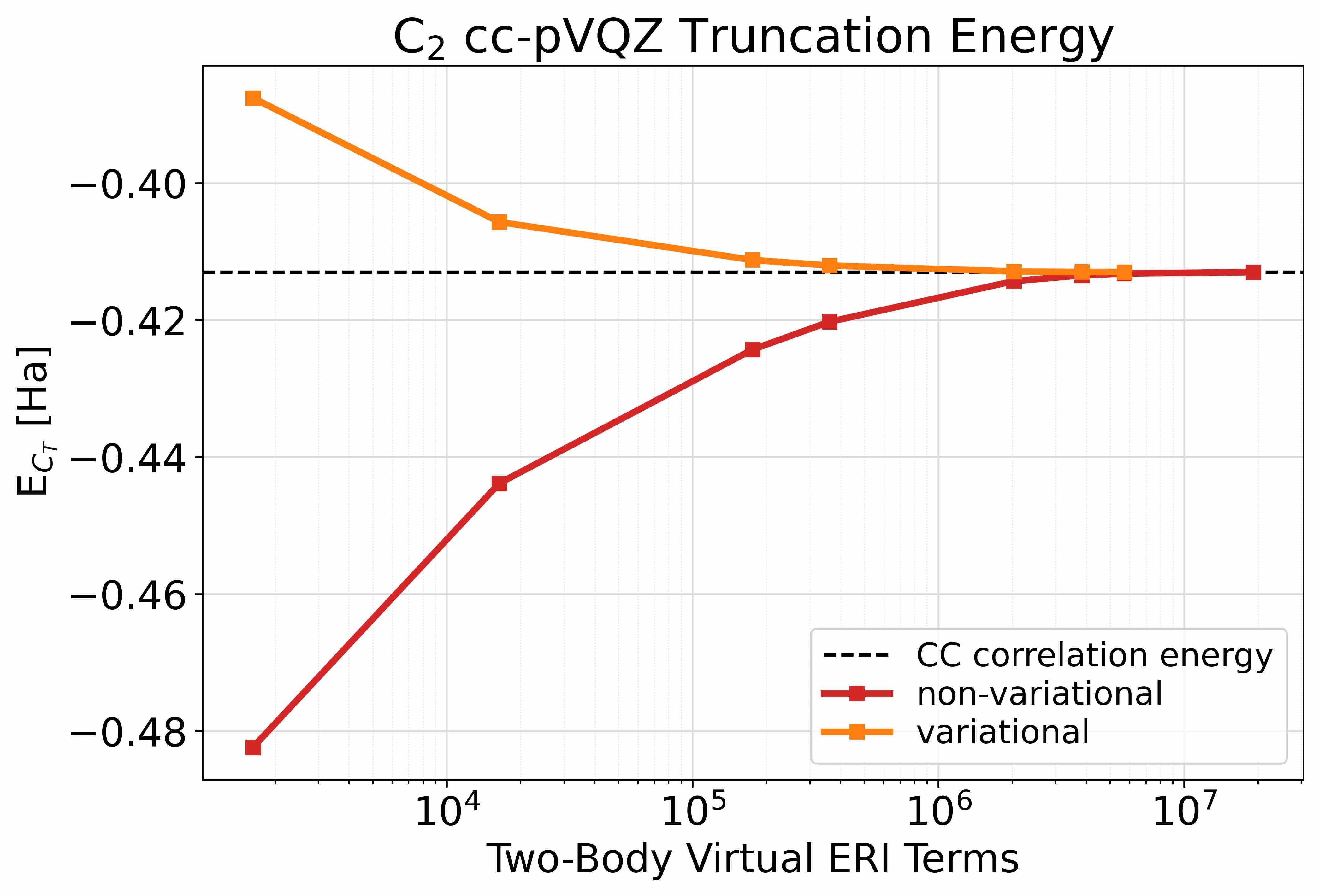}\hfill
    \includegraphics[width=.24\textwidth]{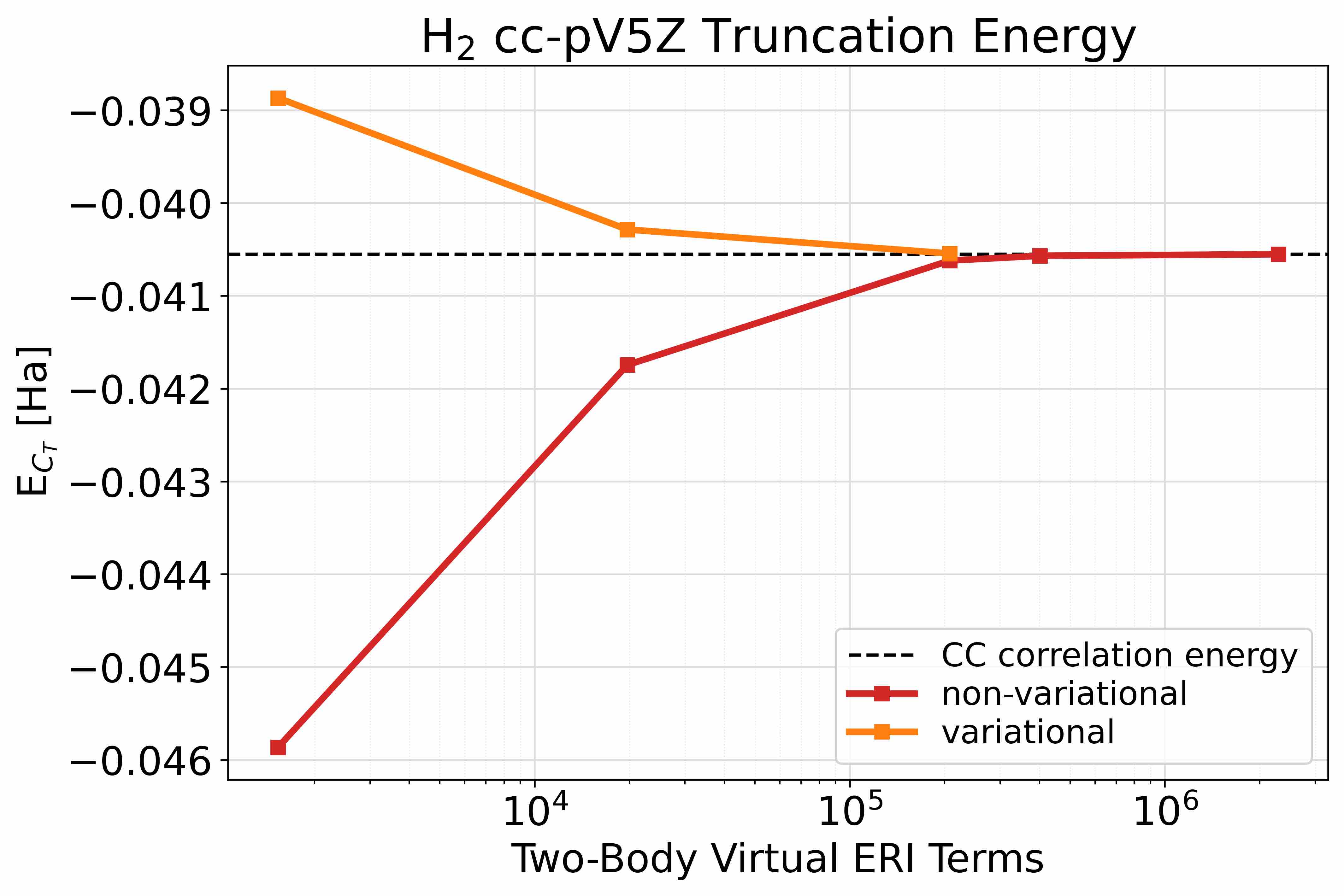}\hfill
    \includegraphics[width=.24\textwidth]{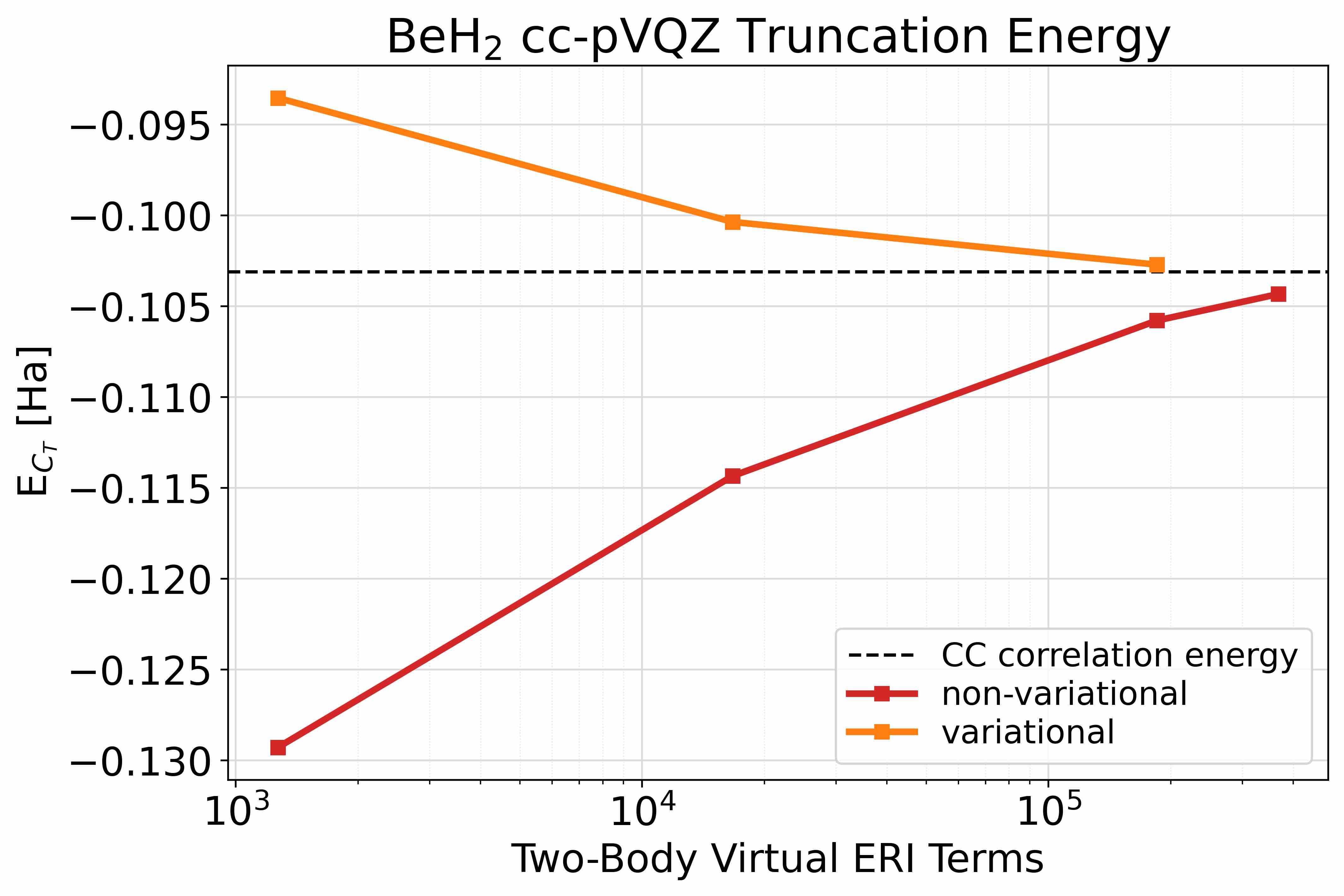}\hfill
    \caption{The semi-logarithmic plots of the number of $vvvv$ terms added into the truncated Hamiltonian versus the approximate coupled cluster correlation energy after applying variational and non-variational SQuISH using to the \ce{NH3}, \ce{HF}, \ce{LiH}, \ce{F2}, \ce{C2}, \ce{H2}, and \ce{BeH2} Hamiltonians using coupled cluster.}\label{fig:cc_allmolecule_energies}
\end{figure}

\begin{figure}[H]
    \includegraphics[width=.24\textwidth]{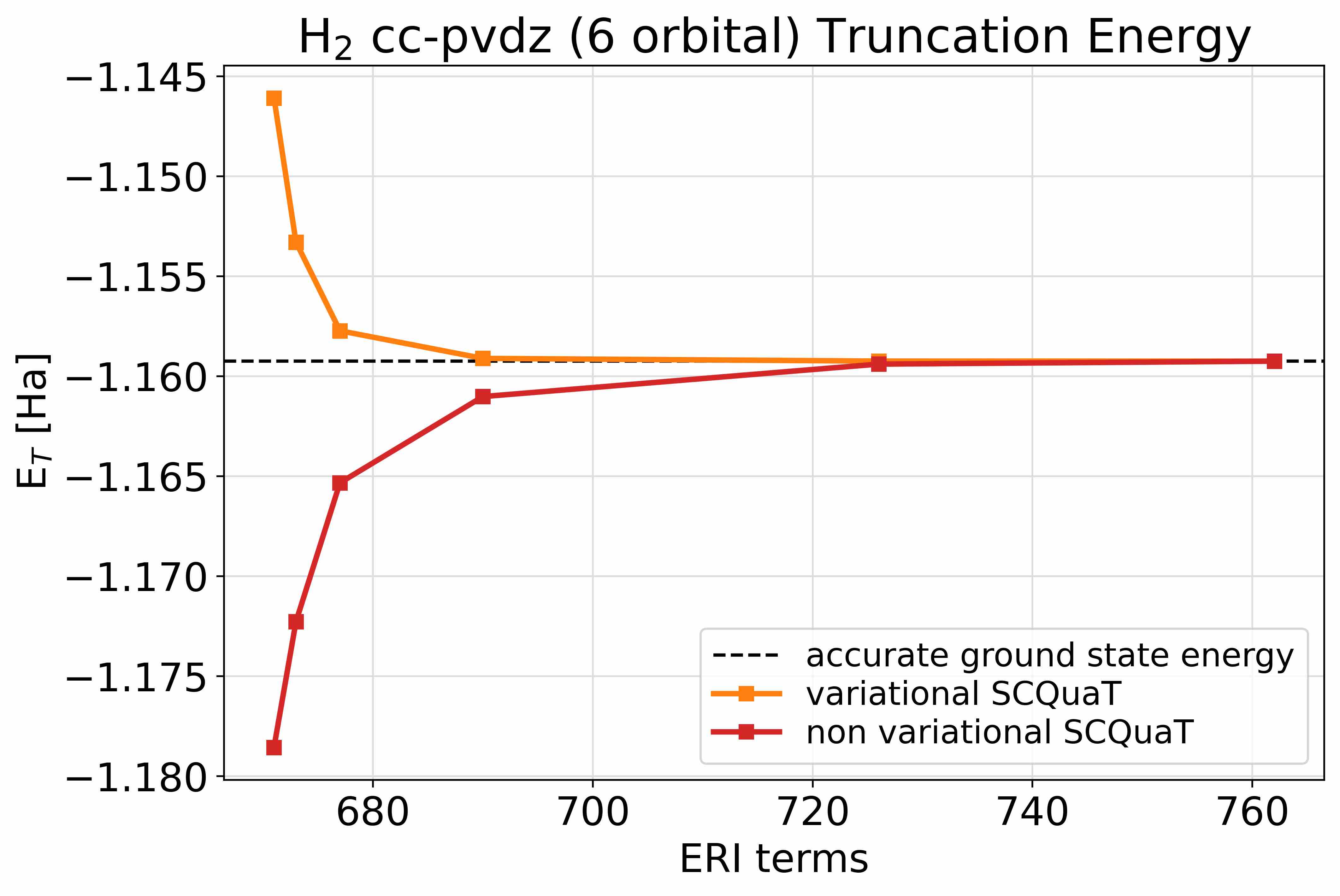}\hfill
    \includegraphics[width=.24\textwidth]{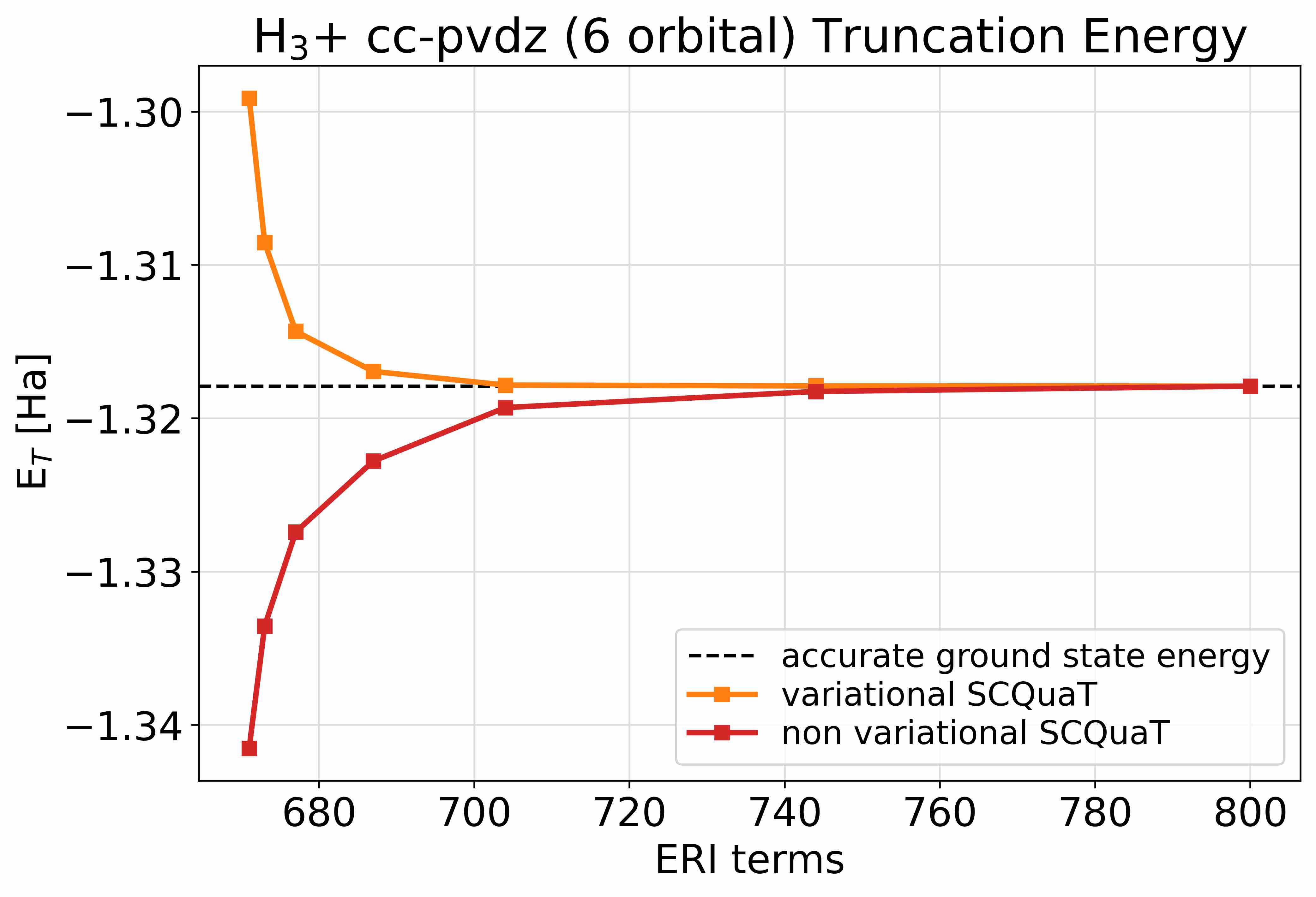}\hfill
    \caption{The plots of the number of $vvvv$ terms added into the truncated Hamiltonian versus the approximate energy after applying variational and non-variational SQuISH for the \ce{H2} and \ce{H3+} Hamiltonians using configuration interaction.}\label{fig:ci_allmolecule_energies}
\end{figure}

\begin{figure}[H]
    \includegraphics[width=.24\textwidth]{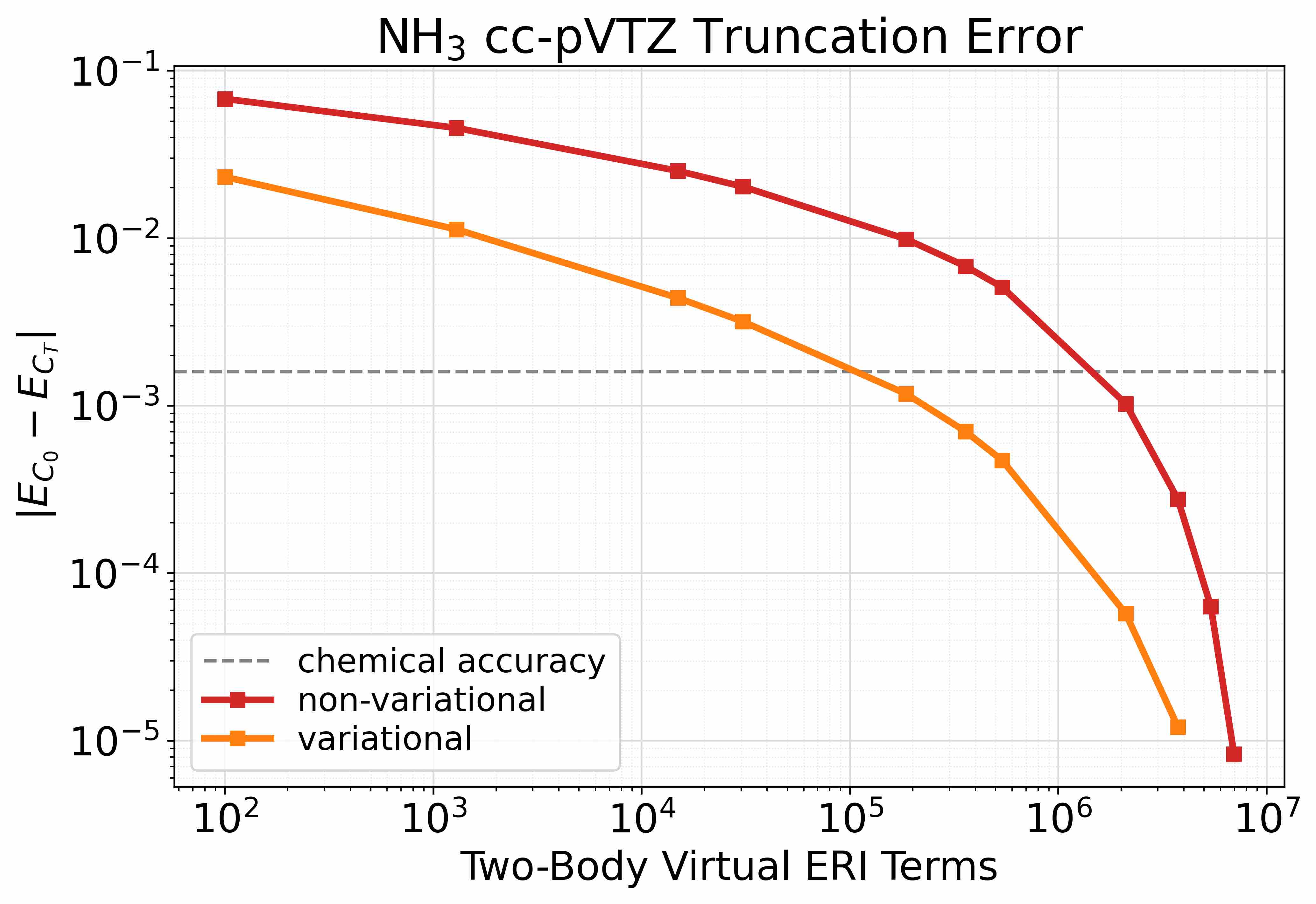}\hfill
    \includegraphics[width=.24\textwidth]{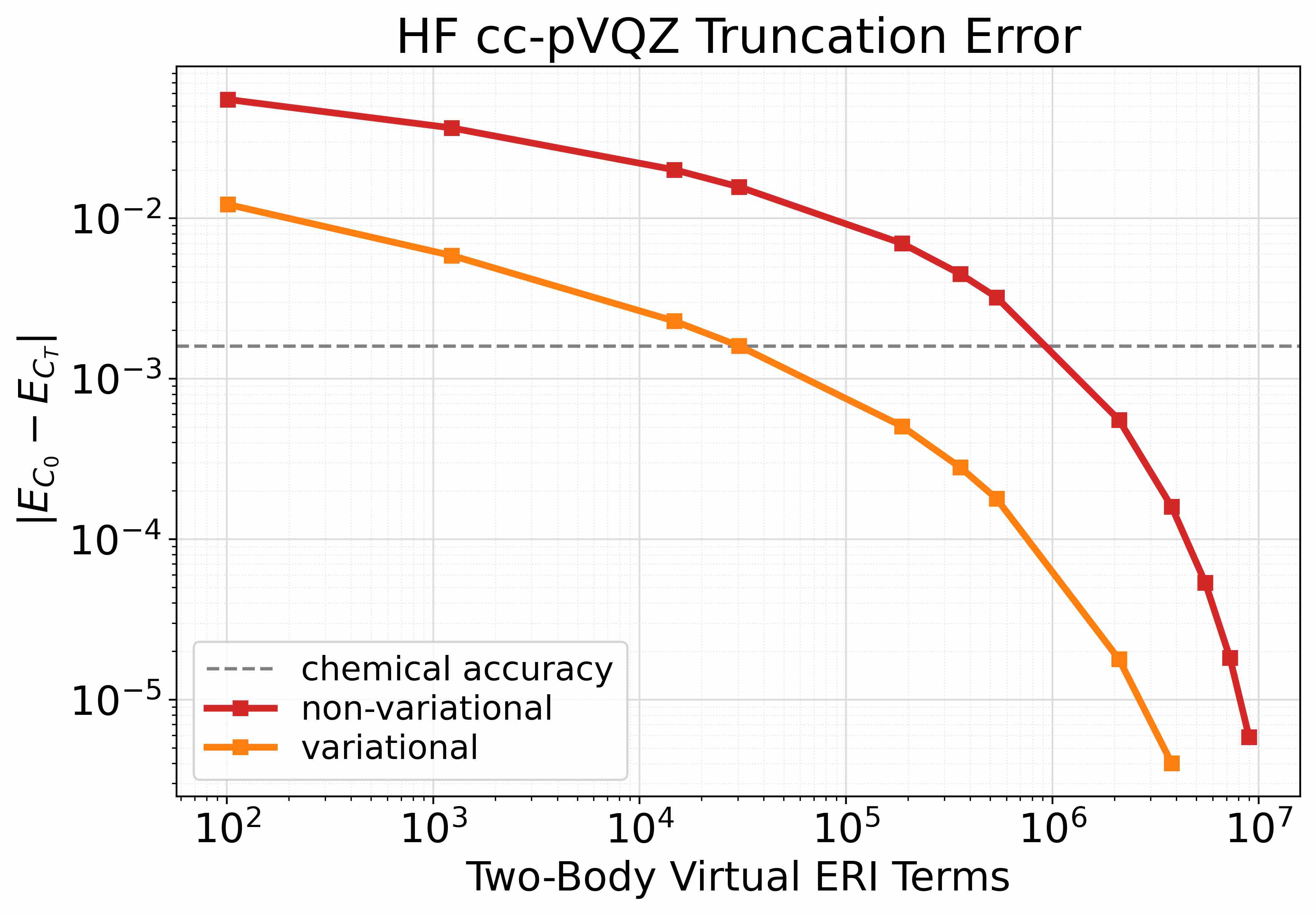}\hfill
    \includegraphics[width=.24\textwidth]{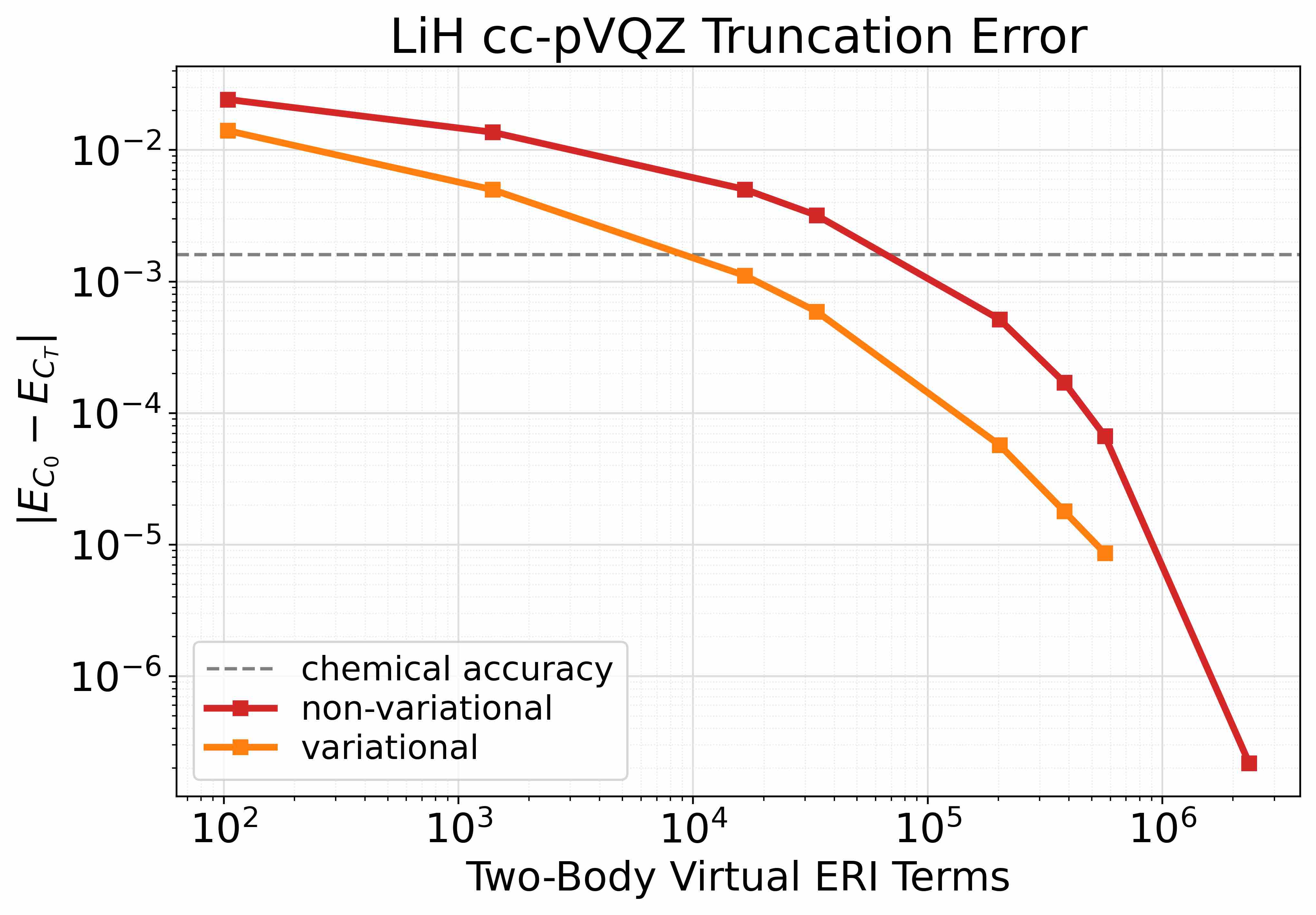}\hfill
    \includegraphics[width=.24\textwidth]{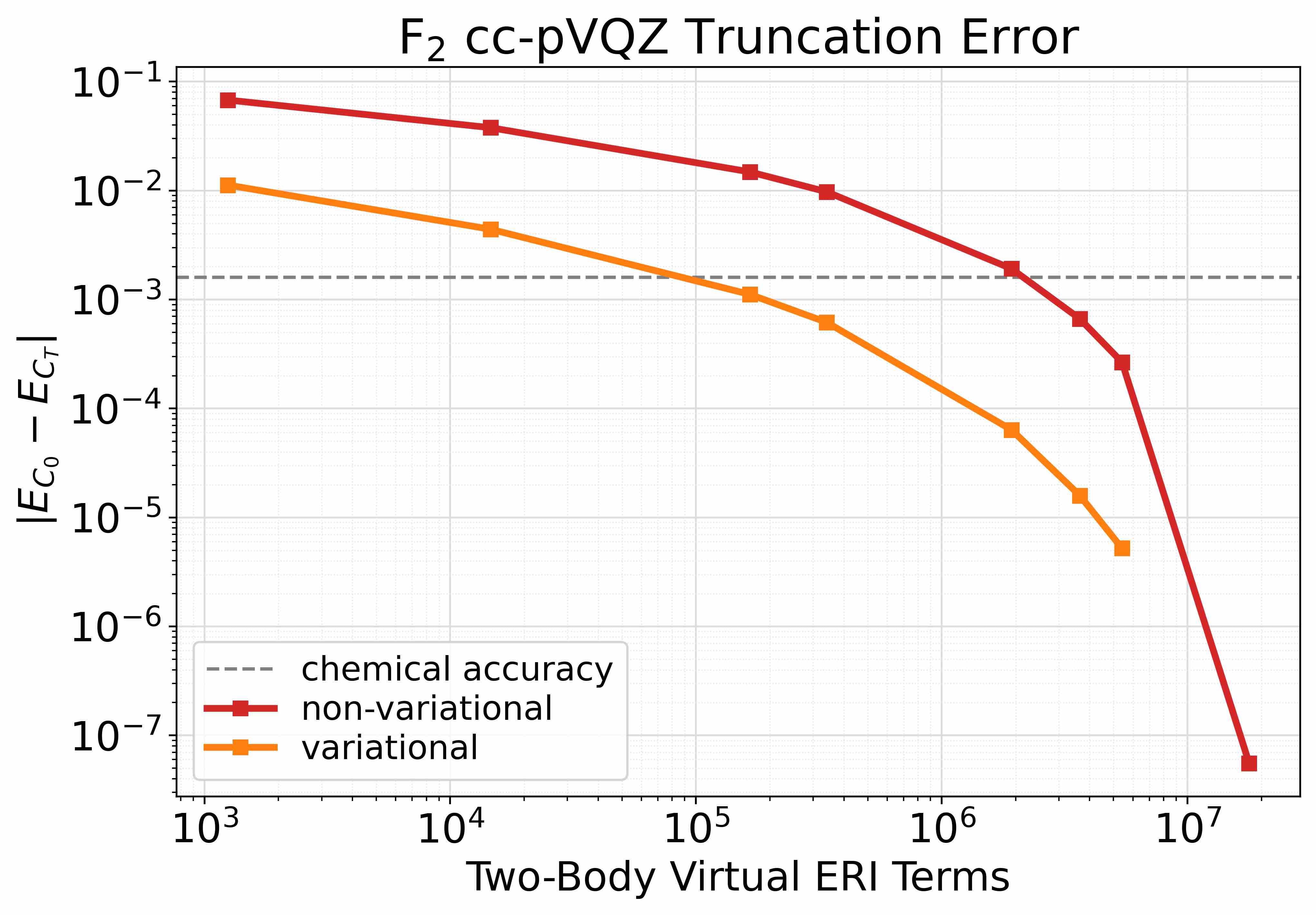}
    \\[\smallskipamount]
    \includegraphics[width=.24\textwidth]{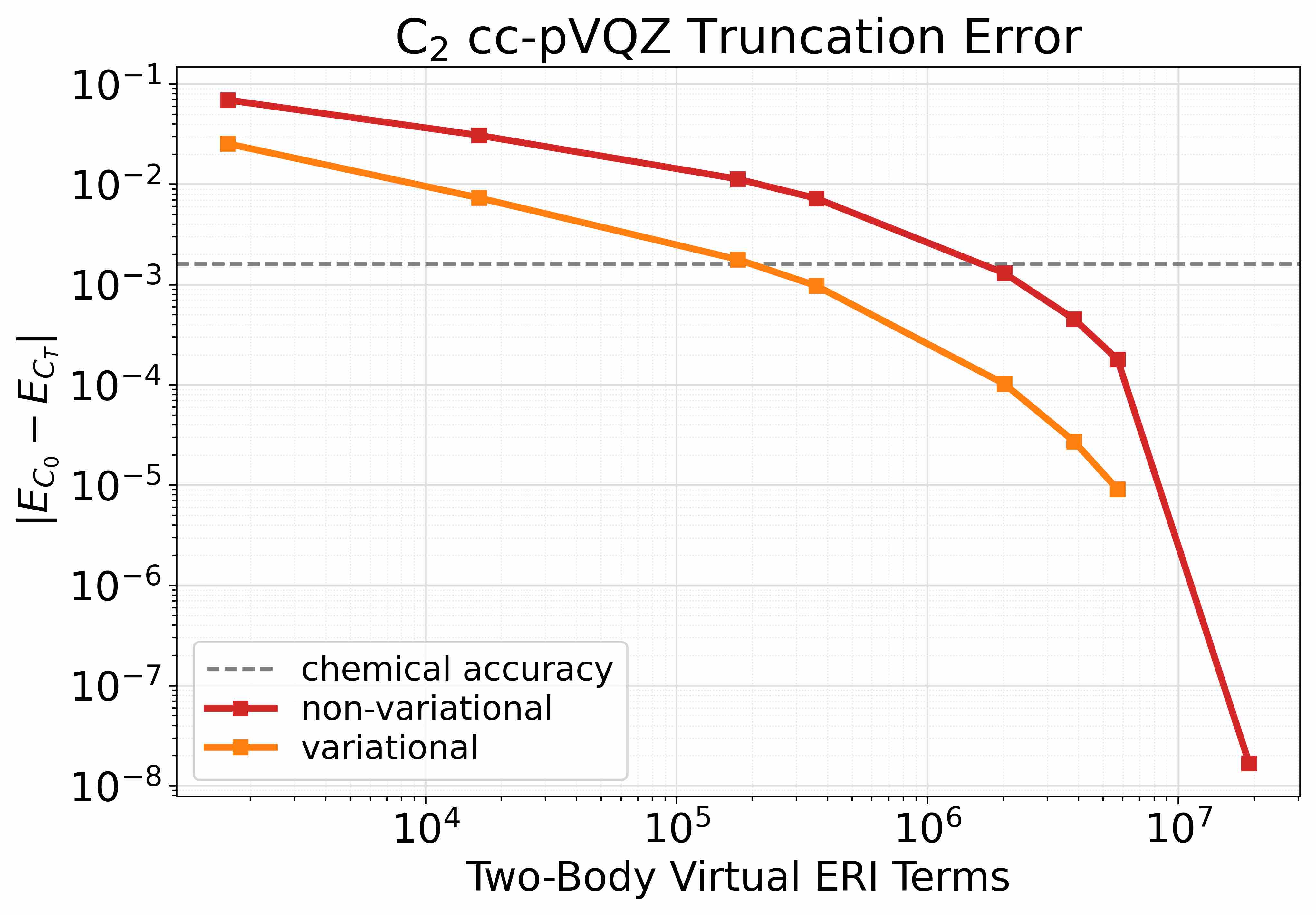}\hfill
    \includegraphics[width=.24\textwidth]{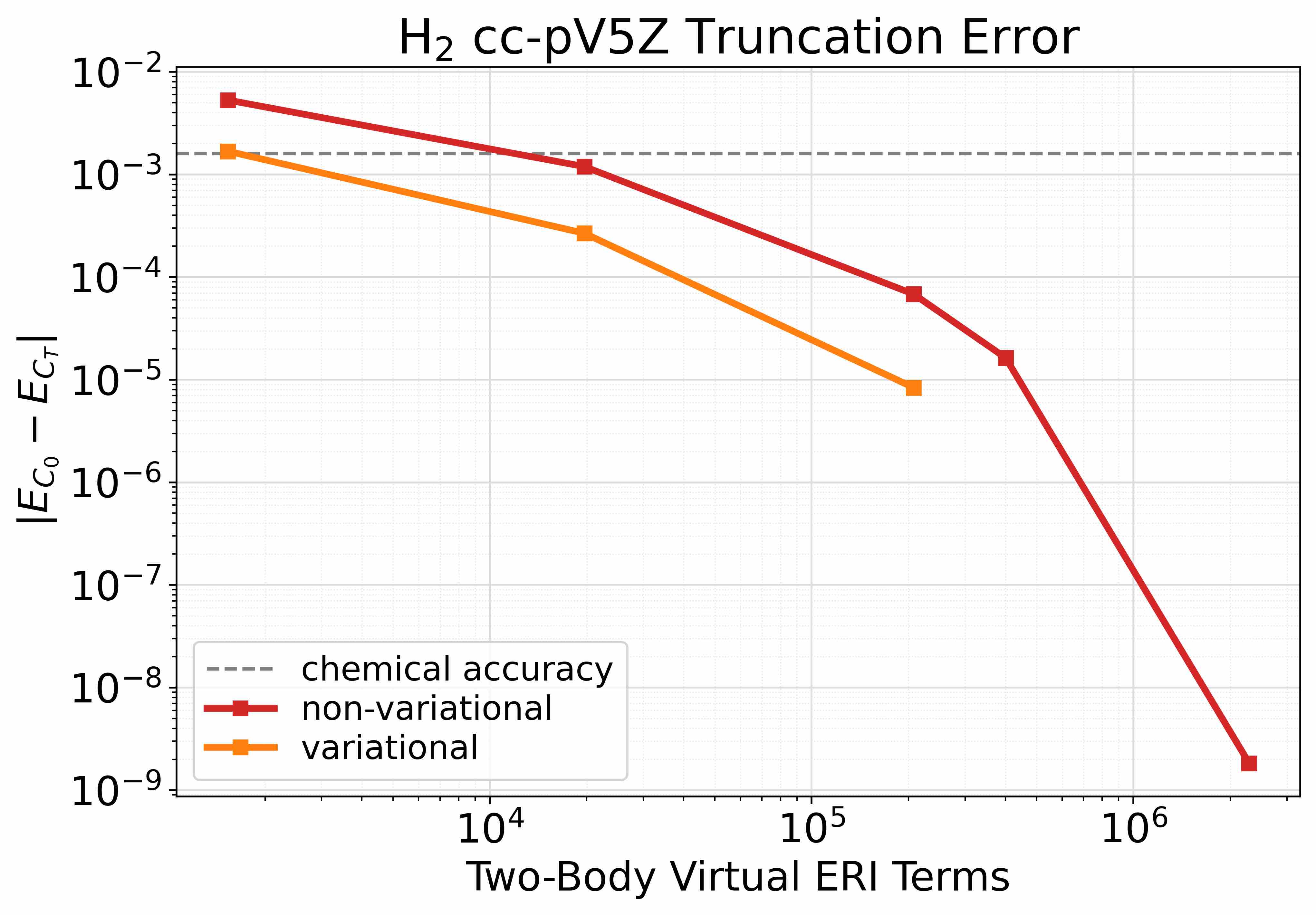}\hfill
    \includegraphics[width=.24\textwidth]{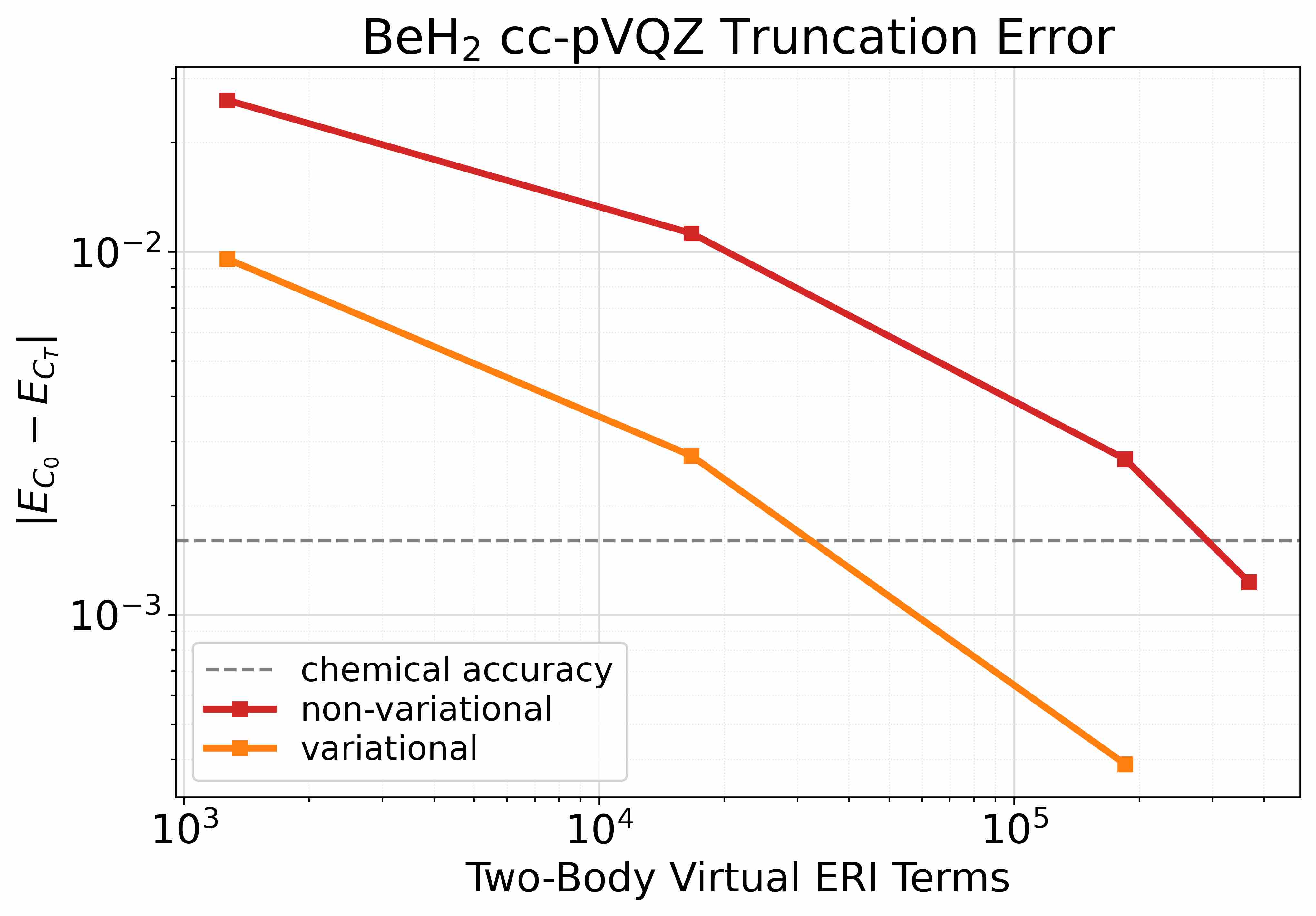}\hfill
    \caption{The plots of the number of $vvvv$ terms added into the truncated Hamiltonian versus the absolute value of the error after applying variational and non-variational SQuISH to the \ce{NH3}, \ce{HF}, \ce{LiH}, \ce{F2}, \ce{C2}, \ce{H2}, and \ce{BeH2} Hamiltonians using coupled cluster.}\label{fig:cc_allmolecule_errors}
\end{figure}

\begin{figure}[H]
    \includegraphics[width=.24\textwidth]{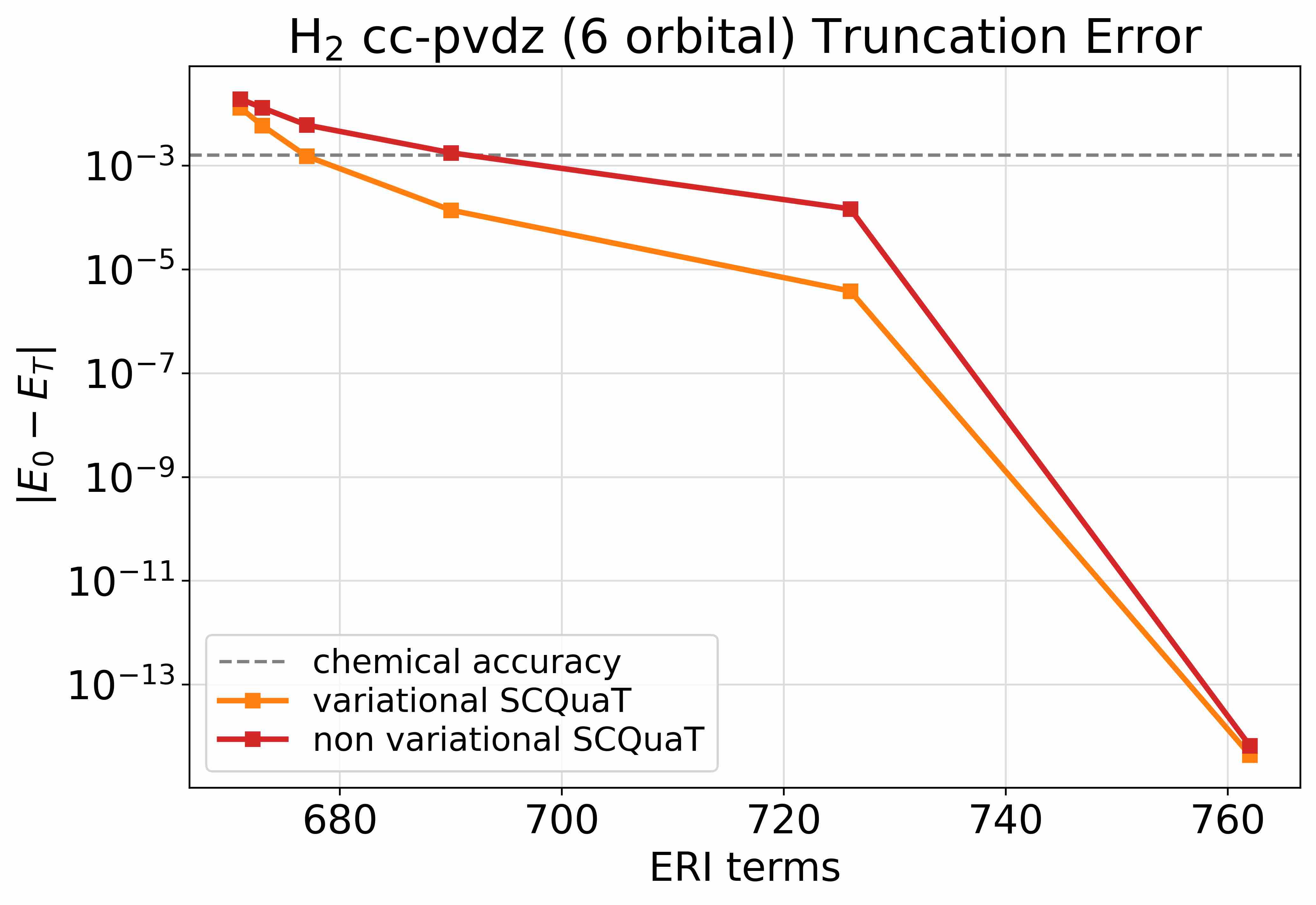}\hfill
    \includegraphics[width=.24\textwidth]{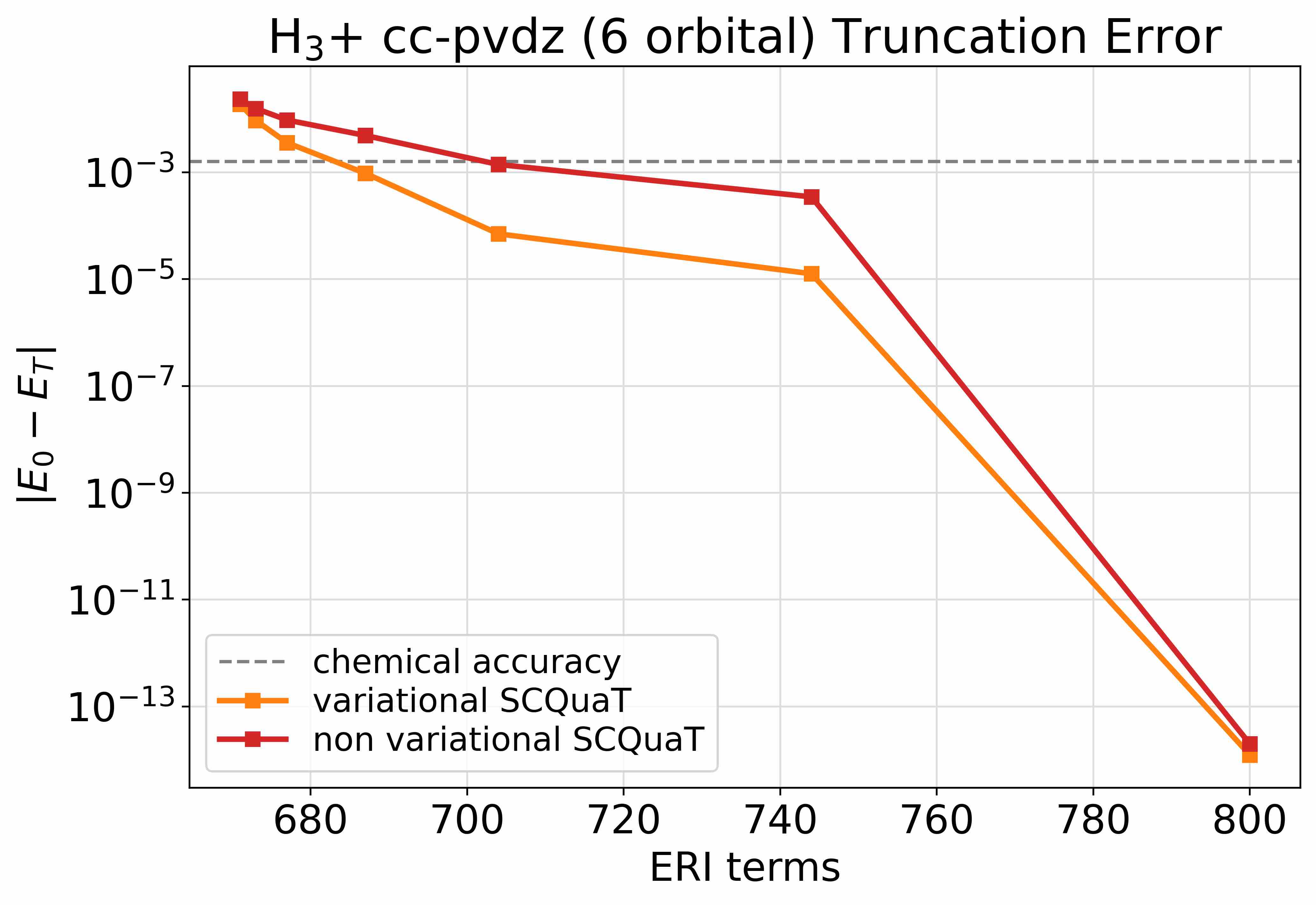}\hfill
    \caption{The plots of the number of $vvvv$ terms added into the truncated Hamiltonian versus the absolute value of the error after applying variational and non-variational SQuISH for the \ce{H2} and \ce{H3+} Hamiltonians using configuration interaction}\label{fig:ci_allmolecule_errors}
\end{figure}

Here show the plots for each of the molecules shown in Table~\ref{tab:CCtable} and Table~\ref{tab:FCItable}.
Fig.~\ref{fig:cc_allmolecule_energies} and Fig.~\ref{fig:ci_allmolecule_energies} are the energy plots for the coupled cluster and configuration interaction runs associated with Table~\ref{tab:CCtable} and Table~\ref{tab:FCItable}, respectively. Fig.~\ref{fig:cc_allmolecule_errors} and Fig.~\ref{fig:ci_allmolecule_errors} are the corresponding error plots.
We terminated SQuISH once the approximate ground state energy was within \num{1.5e-5} and \num{1e-12} of the exact energy for coupled cluster and configuration interaction, respectively.

\begin{figure}[H]
    \centering
    \includegraphics[width=\linewidth]{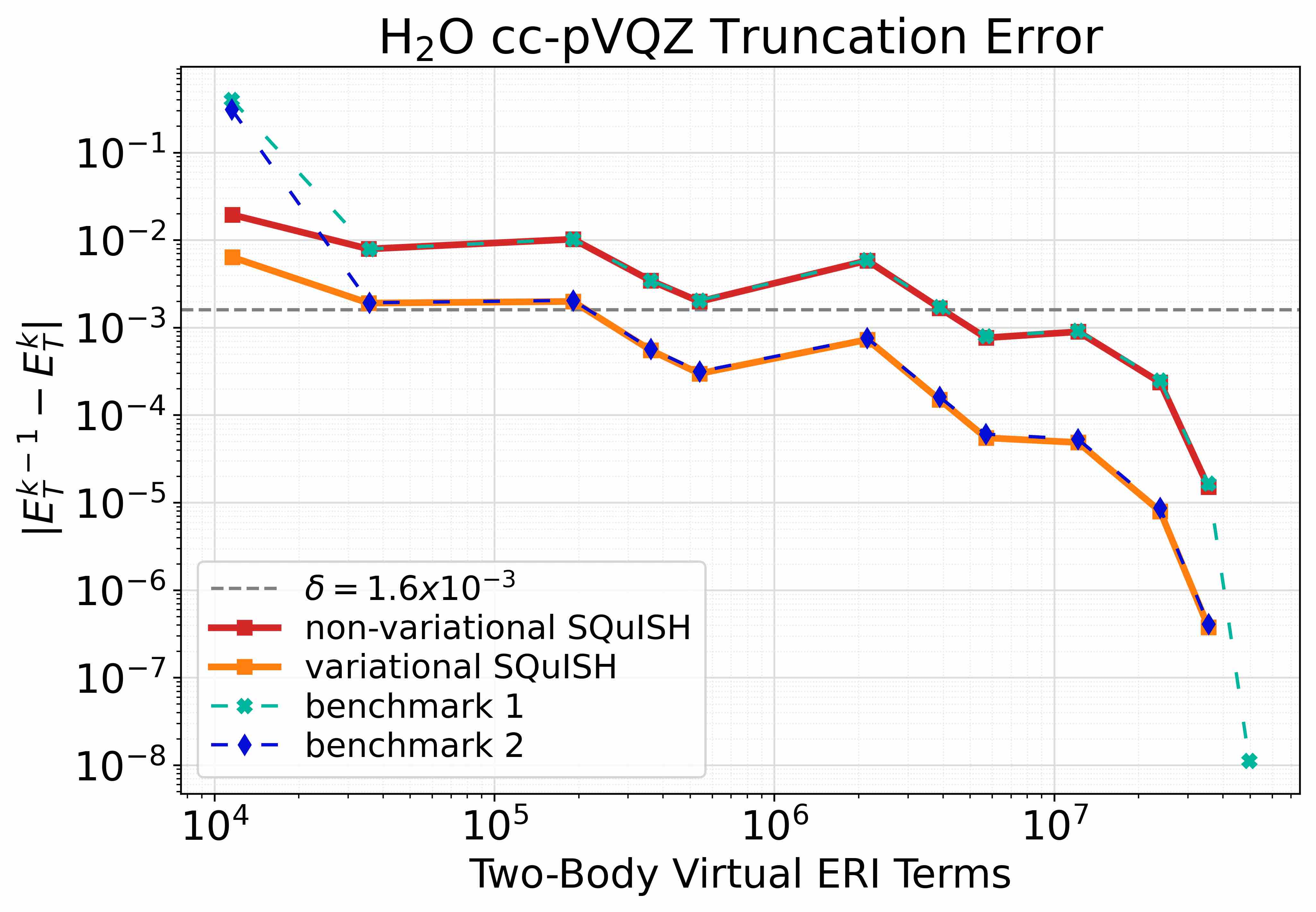}
    \caption{Error plot for the two benchmarks, variational SQuISH, and non-variational SQuISH applied to the \ce{H2O} ccp-VQZ Hamiltonian using $\abs{E_T^{k-1} - E_T^k}$.}
    \label{fig:h20_error2}
\end{figure}

The results presented in the main paper check for convergence in Step 3 using the exact ground state energy to better understand how well the algorithm performs. 
This means that SQuISH terminates when $\abs{E_0 - E_T^k}$ is less than the desired accuracy. 
However, in practice, we are interested in looking at systems to find $E_0$ within some error and do not have access to the exact energy. 
One would employ methods widely used for iterative algorithms. 
In particular, SQuISH checks for convergence using $\abs{E_T^{k-1} - E_T^k}<\delta$.
Fig.~\ref{fig:h20_error2} illustrates the results of using $\abs{E_T^{k-1} - E_T^k}$ to check for accuracy in Step 3 when truncating the \ce{H2O} cc-pVQZ Hamiltonian using SQuISH. 
We note that the error is not monotonic.

\end{document}